\documentclass[AMA,STIX1COL]{WileyNJD-v2}
\usepackage{dsfont}
\usepackage[utf8]{inputenc}
\usepackage{amsthm,amsmath}
\usepackage{bm}
\usepackage{bbm}
\usepackage{moreverb}
\usepackage{booktabs}
\usepackage{tikz}
\usepackage{subfig}
\usepackage{pgfplots}
\usepackage{xcolor}
\usepackage{logicpuzzle}
\usepackage{pdflscape}
\usepackage{csquotes}

\pgfplotsset{every axis/.append style={tick label style={/pgf/number format/fixed},font=\scriptsize,ylabel near ticks,xlabel near ticks,grid=major}}

\articletype{Research Article}

\received{26 April 2016}
\revised{6 June 2016}
\accepted{6 June 2016}

\usepackage{url}
\usepackage{smartdiagram}

\makeatletter
\NewDocumentCommand{\smartdiagramx}{r[] m m}{%
    \StrCut{#1}{:}\diagramtype\option
    \IfStrEq{\diagramtype}{priority descriptive diagram}{% true-priority descriptive diagram
        \pgfmathparse{subtract(\sm@core@priorityarrowwidth,\sm@core@priorityarrowheadextend)}
        \pgfmathsetmacro\sm@core@priorityticksize{\pgfmathresult/2}
        \pgfmathsetmacro\arrowtickxshift{(\sm@core@priorityarrowwidth-\sm@core@priorityticksize)/2}
        \begin{tikzpicture}[every node/.style={align=center,let hypenation}]
        \foreach \smitem [count=\xi] in {#2}{\global\let\maxsmitem\xi}
        \foreach \smitem [count=\xi] in {#2}{%
            \edef\col{\@nameuse{color@\xi}}
            \node[description,drop shadow](module\xi)
            at (0,0+\xi*\sm@core@descriptiveitemsysep) {\smitem};
            \draw[line width=\sm@core@prioritytick,\col]
            ([xshift=-\arrowtickxshift pt]module\xi.base west)--
            ($([xshift=-\arrowtickxshift pt]module\xi.base west)-(\sm@core@priorityticksize pt,0)$);
        }%
        \coordinate (A) at (module1);
        \coordinate (B) at (module\maxsmitem);
        \CalcHeight(A,B){heightmodules}
        \pgfmathadd{\heightmodules}{\sm@core@priorityarrowheightadvance}
        \pgfmathsetmacro{\distancemodules}{\pgfmathresult}
        \pgfmathsetmacro\arrowxshift{\sm@core@priorityarrowwidth/2}
        \begin{pgfonlayer}{background}
        \node[priority arrow,rotate=180,transform shape] (pr-arrow) at ([xshift=-\arrowxshift pt]module\maxsmitem.north west){};
        \end{pgfonlayer}
        \node[below] at (pr-arrow.tip){#3};
        \end{tikzpicture}
    }{}% end-priority descriptive diagram
}%

\pgfplotsset{compat=1.18}

\begin{document}

% \title{A framework for estimands in meta-analyses of clinical trials and a case study on semaglutide versus dulaglutide for Type 2 diabetes}

\title{Incorporating estimands into meta-analyses of clinical trials}

\author[1,0]{Antonio Remiro-Az\'ocar}

\author[2]{Pepa Polavieja}

\author[3]{Emmanuelle Boutmy}

\author[4]{Alessandro Ghiretti}

\author[5] {Lise Lotte Nystrup Husemoen}

\author[6] {Khadija Rerhou Rantell}

\author[13]{Tatsiana Vaitsiakhovich}

\author[7]{David M. Phillippo}

\author[8,9] {Jay J. H. Park}

\author[10]{Helle Lynggaard}

\author[12]{Robert Bauer}

% \author[13]{Frank Kleinjung}

\author[14]{Antonia Morga}

\authormark{REMIRO-AZ\'OCAR \textsc{et al}}

\address[1]{\orgdiv{External Collaboration and Experimentation}, \orgname{Novo Nordisk Pharma}, \orgaddress{\state{Madrid}, \country{Spain}}}

\address[0]{\orgdiv{Department of Statistical Science}, \orgname{University College London}, \orgaddress{\state{London}, \country{United Kingdom}}}

\address[2]{\orgdiv{Launch Evidence Generation and Orchestration}, \orgname{Novo Nordisk Pharma}, \orgaddress{\state{Madrid}, \country{Spain}}}

\address[3]{\orgdiv{Global Epidemiology}, \orgname{Merck Healthcare, KGaA},  \orgaddress{\state{Darmstadt}, \country{Germany}}}

\address[4]{\orgdiv{Data and Statistical Sciences Centre for RWE and EG}, \orgname{Daiichi Sankyo Italia}, \orgaddress{\state{Rome}, \country{Italy}}}

\address[5]{\orgdiv{Real-World Evidence}, \orgname{Novo Nordisk A/S}, \orgaddress{\state{S{\o}borg}, \country{Denmark}}}

\address[6]{\orgname{Medicines and Healthcare products Regulatory Agency}, \orgaddress{\state{London}, \country{United Kingdom}}}

\address[13]{\orgdiv{General Medicines GSA}, \orgname{Sanofi-Aventis Deutschland GmbH}, \orgaddress{\state{Berlin}, \country{Germany}}}

\address[7]{\orgdiv{Bristol Medical School (Population Health Sciences)}, \orgname{University of Bristol}, \orgaddress{\state{Bristol}, \country{United Kingdom}}}

\address[8]{\orgname{Core Clinical Sciences Inc.}, \orgaddress{\state{Vancouver}, \country{Canada}}}

\address[9]{\orgdiv{Department of Health Research Methodology, Evidence, and Impact}, \orgname{McMaster University}, \orgaddress{\state{Hamilton}, \country{Canada}}}

\address[10]{\orgdiv{Biostatistics Methods}, \orgname{Novo Nordisk A/S}, \orgaddress{\state{S\o borg}, \country{Denmark}}}

\address[12]{\orgdiv{Launch Evidence Generation and Orchestration}, \orgname{Novo Nordisk A/S}, \orgaddress{\state{S\o borg}, \country{Denmark}}}

\address[14]{\orgdiv{Global Medical Affairs}, \orgname{Astellas Pharma Europe Ltd}, \orgaddress{\state{Addlestone}, \country{United Kingdom}}}

\corres{Antonio Remiro-Az\'ocar, External Collaboration and Experimentation, Novo Nordisk Pharma, Madrid, Spain. \email{aazw@novonordisk.com}. Tel: (+34) 91 334 9800} 

% \presentaddress{Antonio Remiro-Az\'ocar, External Collaboration and Experimentation, Novo Nordisk Pharma, Calle V\'ia de los Poblados 3, Madrid, 28033, Spain}

\abstract{The estimand framework is increasingly established to pose research questions in confirmatory clinical trials. In evidence synthesis, the uptake of estimands has been modest, and the PICO (Population, Intervention, Comparator, Outcome) framework is more often applied. While PICOs and estimands have overlapping elements, the estimand framework explicitly considers different strategies for intercurrent events. We propose a pragmatic framework for the use of estimands in meta-analyses of clinical trials, highlighting the value of estimands to systematically identify and mitigate key sources of quantitative heterogeneity, and to enhance the applicability or external validity of pooled estimates. Focus is placed on the role of strategies for intercurrent events, within the specific context of meta-analyses for health technology assessment. We apply the estimand framework to a network meta-analysis of clinical trials, comparing the efficacy of semaglutide versus dulaglutide in type 2 diabetes. We explore the impact of a treatment policy strategy for treatment discontinuation or initiation of rescue medication versus a hypothetical strategy for the corresponding intercurrent events. The specification of different target estimands at the meta-analytical level allows us to be explicit about the source of heterogeneity, the intercurrent event strategy, driving any potential differences in results. We advocate for the integration of estimands into the planning of meta-analyses, while acknowledging that potential challenges exist in the absence of subject-level data. Estimands can complement PICOs to strengthen communication between stakeholders about what evidence syntheses seek to demonstrate, and to ensure that the generated evidence is maximally relevant to healthcare decision-makers.}


\maketitle

\renewcommand{\thefootnote}{\alph{footnote}}

\newcommand{\specialcell}[2][c]{%
  \begin{tabular}[#1]{@{}l@{}}#2\end{tabular}}

\subsection*{Highlights}

\paragraph{What is already known?}

\begin{itemize}
\item Estimands are increasingly used to formulate research questions in confirmatory clinical trials and throughout the life-cycle of pharmaceutical research. 
\item In evidence synthesis, the PICO (Population, Intervention, Comparator, Outcome) framework is typically used to pose clinical research questions, and the uptake of estimands has been modest. 
\item While PICOs and estimands have overlapping attributes, the estimand framework additionally places emphasis on the relevance of the population-level summary measure and different intercurrent event strategies. 
\end{itemize}

\paragraph{What is new?}

\begin{itemize}
\item We propose a pragmatic framework for the application of estimands to meta-analyses of clinical trials, within the specific context of health technology assessment.  
\item We apply the estimand framework to a network meta-analysis comparing the efficacy of semaglutide versus dulaglutide in type 2 diabetes, exploring the impact of different intercurrent event strategies. 
\item We highlight the value of estimands to systematically address quantitative heterogeneity between trials, and to maximize the applicability of pooled estimates relative to healthcare decision-making contexts. 
\end{itemize}

\paragraph{Potential impact for RSM readers outside the authors’ field}

\begin{itemize}
\item We highlight the relevance of different intercurrent event strategies to the conduct of meta-analyses.
\item We advocate for the integration of estimands, as a complement to PICOs, into the planning of meta-analyses. 
\item Estimands can ensure that evidence syntheses are maximally relevant to healthcare decision-makers, while strengthening communication between stakeholders about what evidence syntheses seek to demonstrate.
\end{itemize}

\section{Introduction}\label{sec1}

There has been a shift from broad to more specific research questions in the design and analysis of confirmatory clinical trials, stimulated by the recent introduction of the estimand framework.\cite{kahan2024estimands} Estimands have been gradually adopted by regulatory agencies worldwide,\cite{ICH-implementation} are now fundamental in the protocols and statistical analysis plans of Phase 3 clinical trials, and have been incorporated into guidelines such as CONSORT and SPIRIT,\cite{kahan2023reporting, kahan2023consensus} which suggests that the reporting of estimands will become standard practice in clinical trial scientific publications. Estimands now encompass almost the entire life cycle of pharmaceutical research, with applications from early clinical development\cite{homer2022early, englert2023defining, lynggaard2025applying} through to observational and post-authorization safety studies.\cite{ma2025review, li2022estimands, luijken2023tell} 

A similar transition from broad to more narrow research questions has taken place in evidence synthesis for healthcare decision-making. In health technology assessment (HTA), evidence synthesis methods are used to inform policy and reimbursement decisions, made for specific clinical settings with well-defined target populations.\cite{ades2024twenty, remiro2024broad} Given this narrow perspective of research questions, evidence synthesis faces two important challenges: (1) quantitative heterogeneity between studies; that is, variation in treatment effects due to trials addressing different research questions, with potential misalignments in patient populations, treatment implementations and outcomes;\cite{ades2024twenty} and (2) the, often ambiguous, applicability or external validity of pooled treatment effect estimates with respect to the decision-making context.\cite{manski2020toward, sobel2017causal, dahabreh2020toward, rott2024causally} 

The term ``external validity'' often refers to the extent to which inferences, drawn from a specific analysis, apply to a target population.\cite{degtiar2023review, findley2021external} In this article, we take a more general view of external validity or ``applicability'' -- defining these terms as the extent to which inferences apply to the decision-making context (in the HTA case, routine clinical practice in the corresponding jurisdiction). Namely, we do not exclusively relate the terms to the \textit{population} element of research questions,\cite{egami2023elements} but also to other dimensions corresponding to attributes of estimands. In particular, we focus on intercurrent events and intercurrent event strategies, at the heart of the estimand framework. 

To produce results that are most relevant for HTA decision-making, every effort should be made to: (1) mitigate quantitative heterogeneity between the studies in the evidence base;\cite{ades2024twenty} and (2) optimize the external validity of pooled estimates with respect to the decision-making context.\cite{dahabreh2023efficient} The estimand framework holds promise to address these challenges. Firstly, it aligns with a ``narrow'' perspective of research questions, aiming to reduce the heterogeneity stemming from ambiguous questions. Secondly, it is a language that is used in clinical trial design and reporting, and will become increasingly prevalent in publications informing future evidence synthesis work.  

In this article, we discuss the application of the estimand framework to evidence synthesis, focusing on the specific context of HTA. The following base-case scenario is considered. A pairwise or network meta-analysis, combining the results of several randomized controlled trials (RCTs), is required for HTA. Each RCT has target estimands of its own and has been designed to establish the relative efficacy, safety and benefit-risk of a novel treatment, thereby supporting regulatory approval. While head-to-head comparisons versus placebo may suffice for regulatory approval, HTA bodies must perform comparisons versus all treatment options in clinical practice. Network meta-analyses of direct and indirect treatment comparisons are typically used for this purpose, as highlighted by methodological guidelines for joint clinical assessments under the new European Union (EU) HTA Regulation.\cite{evisynth}

To exemplify the use of the estimand framework in evidence synthesis, we apply estimands to a network meta-analysis comparing the efficacy of semaglutide versus dulaglutide in patients with type 2 diabetes (T2D).\cite{lingvay2022indirect} This practical example shows how estimands can be used as a complement to the PICO (Population, Intervention, Comparator, Outcome) framework, to systematically identify, explain and address key sources of quantitative heterogeneity, and to enhance external validity, thereby ensuring that the synthesized evidence is maximally relevant to healthcare decision-making. While our example features a network meta-analysis, leveraging both direct and indirect evidence, the principles are also
applicable to pairwise meta-analyses. 

This article results from the work of the Estimands Implementation Working Group (EIWG) sub-team in HTA and Real-World Evidence (RWE). The EIWG was established by the European Federation of Pharmaceutical Industries and Associations (EFPIA) and the European Federation of Statisticians in the Pharmaceutical Industry (EFSPI), upon recognizing practical challenges for the implementation of estimands in clinical trials and throughout the life cycle of medicinal products. As part of the HTA and RWE sub-team, we have focused on the implementation of estimands across late-phase activities. Nevertheless, our findings are also applicable to the regulatory pre-market authorization setting. For instance, the 2025-27 three-year work-plan of the European Medicines Agency Methodology Working Party lists ``guidance on how to align estimand attributes across different trials in the context of a meta-analysis'' as a key priority,\cite{EMAMWP} and (network) meta-analyses of historical trials are often used to determine the non-inferiority margin of confirmatory active controlled trials.\cite{lynggaard2024applying, FDANI}

\section{What is an estimand?}\label{sec2}

The design and analysis of confirmatory clinical trials is guided by principles from the International Council for Harmonization (ICH) of technical requirements for pharmaceuticals on human use. An addendum to the ICH guideline on statistical principles for clinical trials, ICH E9 (R1), published in 2019, proposes the estimand framework as a systematic approach to specify clinical research questions.\cite{ICHEMA} An estimand is a precise definition of the treatment effect that is targeted by a clinical trial, which should align the trial's planning, design, conduct and analysis with the scientific question posed by investigators.\cite{ICHEMA}

According to the final version of ICH E9 (R1), an estimand is defined by five attributes: (1) population, (2) treatment, (3) variable (or endpoint), (4) population-level summary measure (the addendum uses the term ``population-level summary''), and (5) strategies for intercurrent events. An intercurrent event is a post-baseline event that affects either the interpretation or existence of endpoint measurements. Examples of intercurrent events are treatment discontinuation, initiation of rescue medication, and terminal events like death. The estimand framework is explicit in outlining five potential strategies for handling intercurrent events: 
\begin{itemize}
\item \textbf{Treatment policy}: includes any effect resulting from the occurrence of the intercurrent event in the definition of the treatment effect;
\item \textbf{Hypothetical}: envisages a hypothetical scenario in which the intercurrent event would not occur;
\item \textbf{Composite variable}: incorporates the occurrence of the intercurrent event into the endpoint definition;
\item \textbf{While-on-treatment}: considers the treatment effect only prior to the occurrence of the intercurrent event;
\item \textbf{Principal stratum}: restricts the population to a subpopulation in which the intercurrent event would not occur.
\end{itemize}
In this article, we focus on the treatment policy and hypothetical strategies, but see Kahan et al,\cite{kahan2024estimands} Keene et al,\cite{keene2023estimands} and Clark et al\cite{clark2022estimands} for more information about different strategies for handling intercurrent events. 

ICH E9 (R1) was largely motivated by a perceived lack of clarity in the handling of intercurrent events. In the pre-estimands era, analytical approaches denoted as ``intention-to-treat'' (ITT) or ``per-protocol'' handled intercurrent events ambiguously.\cite{keene2023estimands, clark2022estimands} For instance, the ITT principle clearly aligns with a treatment policy strategy and should include all randomized patients under their randomized treatment allocation. ICH E9 (R1) states:\cite{ICHEMA} 

\blockquote{``ICH E9 
introduced the ITT principle in connection with the effect of a treatment policy in a randomised controlled trial, whereby subjects are followed, assessed and analysed irrespective of their compliance to the planned course of treatment (...) Multiple consequences arising from the ITT principle can be  distinguished. Firstly, that the trial analysis should include all subjects relevant for the research question. Secondly, that subjects should be included in the analysis as randomised. Taken directly from the definition of the ITT principle (see ICH E9 Glossary), a third consequence is that subjects should be followed-up and assessed regardless of adherence to the planned course of treatment and that those assessments should be used in the analysis.''}

However, many analyses reported as ITT ``ignored'' intercurrent events by stopping to follow up participants after the occurrence of the events. As such, outcome data for subjects experiencing the intercurrent events would be excluded from the estimation, thereby diverging from the ITT principle. Moreover, treatment effects may have inadvertently been estimated in hypothetical settings where intercurrent events do not take place, e.g., truncating intercurrent events in mixed models for repeated measures or censoring at the intercurrent event in time-to-event settings.\cite{clark2022estimands, mitroiu2022estimation, mitroiu2020narrative} 

ICH E9 (R1) highlights the potential relevance of alternative intercurrent event strategies beyond treatment policy, with Morga et al recently illustrating the relevance of these strategies to HTA decision-making.\cite{morga2023intention} Consider certain cancer trials, where due to perceived ethical reasons, patients randomized to the control group can switch onto the experimental treatment, as rescue medication following disease progression.\cite{manitz2022estimands, latimer2014adjusting} HTA concerns the evaluation of ``real-world'' clinical effectiveness. However, patients under the current standard of care in clinical practice cannot initiate the experimental intervention, because it is not available in the healthcare system. 

As argued by Latimer et al\cite{latimer2014adjusting} and others,\cite{morga2023intention,jackson2025new} a treatment policy strategy that includes any effect resulting from treatment switching as part of the treatment effect definition leads to a treatment pathway that is not relevant for the HTA decision problem. Key to the selection of any estimand is its relevance to stakeholders, and different intercurrent event strategies are likely to serve different stakeholders.\cite{keene2020matters, polverejan2023defining} Similarly, the relevance of an estimand critically depends on the context. Ideally, estimand choice should not be driven by habitual choices of analytical methods and their convenience but by the clinical research question.

A hypothetical strategy for treatment switching, considering the setting where rescue medication is unavailable, would offer greater external validity relative to the HTA decision-making context.\cite{jackson2025new} Nevertheless, its estimation rests on unverifiable assumptions -- when imputing outcomes for the control patients that have switched treatment -- which may undermine the internal validity offered by randomization. The Institute for Quality and Efficiency in Health Care in Germany, has prioritized internal over external validity, and is opposed to diverging beyond treatment policy (and composite variable strategies for terminal events like death, where post-intercurrent event outcomes do not conceptually exist).\cite{morga2023intention} While these strategies are generally considered easier to estimate, note that their estimation may still require strong assumptions in the presence of missing outcomes.\cite{bell2025estimation} 

% It could also lead to ``double-counting'' when applying treatment discontinuation rates to cost-effectiveness models. 

Other HTA agencies, such as the National Institute for Health and Care Excellence (NICE) in England and Wales, acknowledge that departures from treatment policy might be useful.\cite{morga2023intention} This is a view which we share, particularly since HTA demands for external validity, and also since payers assess clinical effectiveness at the level of their jurisdiction. Treatment pathways will differ between countries or regions. A treatment policy strategy where patients switch to a treatment that is unavailable in the local context is arguably of little relevance to HTA decision-makers in such context. 

\section{Estimands for evidence synthesis}\label{sec3}

\subsection{Estimands as complements to PICOs}\label{subsec31}

We focus on the specific context of evidence synthesis for HTA, where the uptake of estimands has been modest, and where the strategy for intercurrent events is often a secondary consideration.\cite{remiro2024broad} For instance, the Cochrane Risk of Bias 2 (RoB 2) tool,\cite{sterne2019rob} recommended by HTA guidance to assess RCT risk-of-bias in evidence syntheses,\cite{validity} does not reflect estimands and is misaligned with the intercurrent event strategies presented by ICH E9 (R1).\cite{poythress2023hta50}

In evidence synthesis and HTA, the PICO (Population, Intervention, Comparator, Outcome) framework is typically used to formulate research questions. PICOs and estimands are closely connected, with the former preceding ICH E9 (R1) by several decades.\cite{richardson1995well} The population is included in both frameworks; the intervention and comparator elements of PICO correspond to the treatment attribute of estimands; and the outcome element of PICO is similar to the variable (or endpoint) attribute of estimands, but often broader.\footnote{Notably, the new EU HTA Regulation Joint Clinical Assessment (JCA) ``Guidance on outcomes for joint clinical assessments'' states that ``outcomes are distinct from the way in which they are measured, and an outcome can be measured in several ways'', and that the ``measure of an outcome'' (i.e., how the outcome is assessed at the subject-level, including use of a specific outcome measure instrument), corresponds to the variable or endpoint estimand attribute in ICH E9 (R1).\cite{outcomes} For instance, ``if the outcome is pain, the measure of the outcome could be, for example, the change in the level of pain on a patient-reported numeric rating scale (from 0-10) at 6 months after initiation of the treatment.''\cite{outcomes}} While strategies for handling intercurrent events are at the heart of ICH E9 (R1), these -- and the population-level summary measure -- are not considered by PICO. All else being equal, PICOs formulate broader research questions than estimands, as estimands contain two additional attributes. 

In this manuscript, we do not consider the use of PICO to formulate research questions at the individual trial level, but to postulate research questions across the synthesis of multiple trials. PICOs are typically used to guide the data extraction process for systematic literature reviews and meta-analyses,\cite{frandsen2020using} and are recommended by the PRISMA statement and its network meta-analysis extension.\cite{moher2010preferred, hutton2016prisma} In HTA, the PICO framework is well-established to determine the scope of assessments.\cite{jen2025current} Notably, the EU Joint Clinical Assessment (JCA) scope is based on sets of PICO questions posed by member states.\cite{van2024impact}  

Within the aforementioned context, the relative imprecision of the PICO framework is useful, and potentially desirable, for pragmatic or feasibility reasons. Estimands are selected by sponsors (in agreement with regulators) that can align their trial's design, data collection and analysis with the desired research question. Conversely, PICOs are generally policy-driven, not necessarily driven by the available data,\cite{scoping} and -- in an evidence synthesis context -- determined after some or most trials have been conducted. As such, ``meta-analysts'' often rely on the analysis of secondary data, typically with limited input on primary data collection and without full access to subject-level data.\cite{remiro2022parametric} 

Moreover, presume that HTA agencies were to ask very narrow research questions, by requesting specific population-level summary measures and intercurrent event strategies as part of the scope. This would produce a sparser evidence base, of fewer trials, potentially resulting in disconnected networks that threaten the feasibility of network meta-analyses. In the EU JCA context, broader research questions can facilitate the consolidation of scopes, particularly given that a wide variability in policy questions across member states is anticipated.\cite{van2024impact} 

Consider the JCA guidance issued by the EU Member State Coordination Group on HTA, ``Guidance on the scoping process'',\cite{scoping} and ``Guidance on outcomes for joint clinical assessments''.\cite{outcomes} We shall refer to these documents as the ``scoping guidance'' and the ``outcomes guidance'', respectively. The scoping guidance states that the population for the assessment scope PICO ``should be as specific as possible and avoid ambiguity''.\cite{scoping} The population is based on the claimed therapeutic indication; typically, the indication in the health technology developer's submission to the European Medicines Agency.\cite{scoping} As such, the population attribute of the estimand submitted to regulators is expected to overlap with that of the PICO in the HTA scope. 

While the population attributes are similarly narrow across frameworks, this is not generally the case for the intervention, comparator and outcome. According to the EU JCA scoping guidance, ``variations of the intervention, such as dose or timing of administration, (...) do not require a separate PICO''.\cite{scoping} Similarly, the treatments comprising a comparator can be drug classes, and ``should not contain information on dosage or regimen''.\cite{scoping} According to the outcomes guidance, the timing of outcomes does not need to be specified as part of the PICO scope, and outcomes can be measured in different ways without necessarily specifying the way in which they are measured.\cite{outcomes} Conversely, estimands should define precisely the dosage, regimen and mode of administration of treatments, the time at which endpoints are measured, and may include the use of specific measure instruments indicating how the endpoint is assessed at the subject level. 

The breadth of research questions in HTA scoping is not limited to the EU, but also apparent in other jurisdictions. Consider the NICE assessment of cabozantinib for previously treated advanced differentiated thyroid cancer, unsuitable for or refractory to radioactive iodine, for which final technology appraisal guidance (TA928) was published in November 2023.\cite{NICEguidance} The evidence from cabozantinib in the HTA submission was largely based on the results of the pivotal Phase 3 RCT COSMIC-311, also used to obtain marketing authorization worldwide.\cite{brose2021cabozantinib} For illustrative purposes, we display below the PICO elements (Table \ref{PICO-cabo}) in the final scope of the NICE technology appraisal of cabozantinib,\cite{NICEscope} and the primary estimands (Table \ref{estimand-cabo}) of the COSMIC-311 trial.\cite{EMAcabometyx}

\begin{table}[!htb]
\centering
\begin{tabular}{|m{11em}||m{33em}|} 
  \hline
  Population & Adults with locally advanced or metastatic differentiated  thyroid carcinoma, whose disease is refractory to, or who are unsuitable for radioactive iodine, and whose disease has progressed during or after prior systemic therapy\\   
  \hline
  Intervention & Cabozantinib \\
  \hline
  Comparator &  Best supportive care \\ 
  \hline
  Outcomes &  
  Overall survival \newline
  Progression-free survival \newline
Response rate \newline
  Adverse effects of treatment \newline
 Health-related quality of life \\
  \hline
\end{tabular}
\caption{PICO elements extracted from the final scope of the NICE technology appraisal of cabozantinib for previously treated differentiated thyroid cancer unsuitable for or refractory to radioactive iodine.
\cite{NICEscope}}
\label{PICO-cabo}
\end{table}

\begin{table}
\centering
\begin{tabular}{|m{6em}||m{20em}|m{20em}|} 
  \hline
  Estimand & Objective response rate & Progression-free survival \\   
  \hline
  Population &  \multicolumn{2}{p{40em}|}{Patients with radioiodine-refractory differentiated thyroid cancer who have progressed after prior VEGFR-targeted therapy} \\
  \hline
  Treatments & \multicolumn{2}{p{40em}|}{Oral cabozantinib (60 mg once daily) \newline Matching placebo}
  \\  
  \hline
   Variable (endpoint) & Radiographic response per RECIST 1.1 & Duration of radiographic progression-free survival \\
  \hline
  Population-level summary measure & Difference in proportions of subjects with a best overall response of confirmed complete response or confirmed partial response per RECIST 1.1 between treatment conditions & Difference in survival functions between treatment conditions \\
  \hline
  Intercurrent event strategies 
  & \textbf{Treatment policy} for receipt of local radiation to bone, surgical resection of non-target tumor lesions, loss to radiographic follow-up, or receipt of local non-protocol anti-cancer medications other than for disease under study 
  \newline
  \textbf{While on treatment} for surgical resection of target tumor lesions, receipt of systemic non-protocol anti-cancer medications, local non-protocol anti-cancer medications for disease under study, or local radiation to soft tissue for disease under study 
  & \textbf{Treatment policy} for clinical deterioration, receipt of local radiation to bone, surgical resection of non-target tumor lesions, or receipt of local non-protocol anti-cancer medications other than for disease under study 
  \newline
  \textbf{Hypothetical} for surgical resection of target tumor lesions, receipt of systemic non-protocol anti-cancer medications, local non-protocol anti-cancer medications for disease under study, or local radiation to soft tissue for disease under study \\
  \hline 
\end{tabular}
\caption{Primary estimands of the COSMIC-311 trial.\cite{EMAcabometyx}}
\label{estimand-cabo}
\end{table}

The population attributes are almost identical across frameworks and are narrowly defined, corresponding to the claimed therapeutic indication. Conversely, while the treatment attribute of the estimand specifies dosage, regimen and mode of administration, the intervention and comparator PICO attributes do not. Similarly, the estimand variables (endpoints) are more specific than the PICO outcomes, which are arguably more ``patient-relevant'',\cite{outcomes} featuring overall survival, adverse effects of treatment, and health-related quality of life. In all cases, the estimand attributes are encompassed by their corresponding PICO elements. While the estimand framework explicitly addresses strategies for intercurrent events and defines a population-level summary measure, the PICO framework does not. The tables should demonstrate the greater precision employed by the estimand framework in specifying the clinical research question. 

\begin{figure}[htb]
\centering\includegraphics[width=\linewidth]{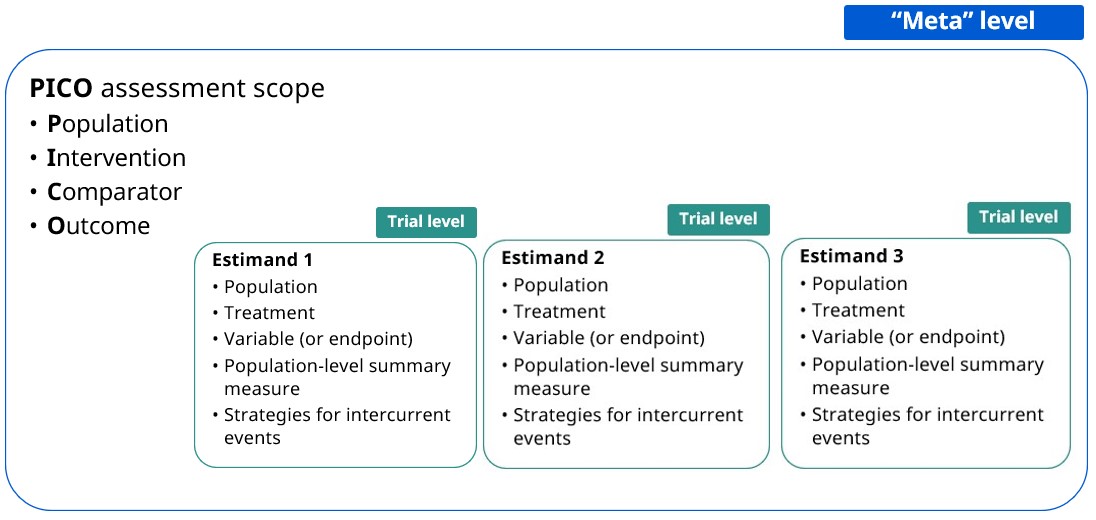}
  \caption{PICOs can be viewed as sets of different estimands, with each PICO encompassing multiple estimand definitions.\cite{remiro2024broad} The ``meta'' level refers to the meta-analytical level, across the trials in the evidence base defined by the assessment scope PICO.}
  \label{PICO-vs-estimands}
\end{figure}

With this article, we do not intend to incriminate the use of PICOs, or to suggest indiscriminately replacing PICOs with estimands. We acknowledge that the breadth of PICOs is attractive for scoping, to describe the ``totality'' of evidence in qualitative evidence syntheses and health technology assessments. As illustrated in Figure \ref{PICO-vs-estimands}, we regard PICOs as sets of different estimands, with a given PICO potentially containing several estimand definitions.\cite{remiro2024broad} Our aim in the next sections is to promote the coexistence of estimands with PICOs and the use of estimands ``as a complement to the PICO framework'', as proposed by the EU JCA outcomes guidance.\cite{outcomes} Our proposal departs from the original idea by Remiro-Az\'ocar of completely replacing the PICO framework with a ``PICOSI'' specifying the additional ``Summary measure'' and ``Intercurrent event strategies'' estimand attributes,\cite{remiro2022some} now recommended by Canada's Drug Agency in all evidence submitted for reimbursement decision-making.\cite{CDA} 

\subsection{Estimands for (network) meta-analysis}\label{subsec32}

The focus of our proposed framework is on pairwise and network meta-analyses, typically used to pool treatment effects across multiple RCTs. It is assumed that the RCTs have been primarily designed to support regulatory approval, each RCT with target estimands of its own. Meta-analyses of direct and indirect treatment comparisons are then required for HTA purposes, as highlighted by JCA methodological guidelines for the new EU HTA Regulation.\cite{evisynth}

There has been limited discussion in the evidence synthesis literature to date about potential implications of the estimand framework for (network) meta-analysis. In our view, the practice of pairwise and network meta-analysis faces two important challenges in healthcare decision-making: (1) quantitative heterogeneity between trials; that is, variation in treatment effects due to trials addressing different research questions;\cite{ades2024twenty} and (2) the often ambiguous external validity or applicability of pooled treatment effect estimates with respect to specific clinical research questions.\cite{manski2020toward, sobel2017causal} We suspect both challenges to be more problematic in the synthesis of trials conducted by different sponsors than in the pooling of trials within the same Phase 3 development program. 

The estimand framework holds promise to address these two challenges, by: (1) identifying, explaining and potentially reducing quantitative heterogeneity between the trials in the evidence base; and (2) optimizing the relevance of pooled estimates with respect to the decision-making context. While it suffices to think of estimands at the individual trial level for the first challenge, the second challenge requires viewing estimands at two different hierarchical levels: (1) the trial level; and (2) the meta-analytical (``meta'') level.\cite{remiro2022some} 

\subsubsection{Aligning estimands at the trial level}\label{subsec321}

The ICH E9 (R1) addendum, while lacking explicit guidelines for estimands in meta-analysis, already warns against a ``na\"ive comparison between data sources, or integration of data from multiple trials without consideration of the estimand that is addressed in each''.\cite{ICHEMA} Individual trials are not necessarily planned with similar estimands, nor in anticipation of a meta-analysis. Almost invariably, there will be some misalignment between estimand components across trials; in patient populations, treatment implementations, endpoint definitions, reported population-level summary measures, intercurrent events, or strategies for intercurrent events.

Critically, discrepancies between estimand attributes may arise even where the high-level estimand descriptions appear to be the same. Patient populations that are seemingly identical, based on estimands' population attribute, can differ due to discrepancies in trial selection criteria. Versions of administered treatment can differ in dosing regimen, delivery mechanism, or concomitant background medications, which may also change over time. Endpoints assessed at similar time points may have been measured using different instruments. Even the seemingly consistent definition of an intercurrent event strategy like treatment policy for treatment switching can give way to heterogeneous treatment regimens across studies, since standard of care and rescue medication are likely to change over time and over trials.

There is a need to align estimand attributes across trials, but this can be challenging, particularly under limited access to subject-level data. A manufacturer submitting evidence to HTA bodies typically has access to subject-level data from at least one ``index'' RCT, comparing the efficacy of its novel active treatment versus a control, but only aggregate-level summary data from ``competitor'' RCTs. In this scenario, the manufacturer is able to align  estimands with respect to the competitor trials, but, awkwardly, not with respect to its own RCT(s).\cite{phillippo2018methods} In any case, what estimands do provide is a framework for transparent and harmonized reporting at the trial level, and for more standardized treatment effect definitions within therapy areas. As illustrated by our case study in Section \ref{sec4}, this is useful to explicitly identify potential misalignments in the evidence base, and to explain key sources of quantitative heterogeneity between trials, using a language that is well-established in trial design and reporting.

\subsubsection{Defining estimands at the meta-analytical level}\label{subsec322}

If estimands impact the interpretation of individual clinical trials, one can conclude that they will impact the interpretation of syntheses of such trials. Behind every trial-level analysis or estimator, there is an estimand being targeted. This is no different for meta-analyses. The pooling of estimates targeting different estimands results in meta-analyses that target neither of the estimands, producing an ambiguous average with dubious applicability.\cite{vuong2024estimands} This raises the question of whether one can define an overarching estimand at the ``meta'' level if there is substantial cross-trial heterogeneity between estimands. 

We propose defining an ``ideal'' meta-analytical estimand, which can be used to restrict the evidence base and perform more targeted (network) meta-analyses, enhancing applicability relative to a specific clinical research question. One can formulate a PICO for the assessment scope, then narrow the evidence base using a target meta-analytical estimand, which more precisely defines the relevant question for healthcare decision-makers. In current practice, such target estimand is sometimes implicitly defined. For instance, where HTA agencies oppose intercurrent event strategies that diverge beyond treatment policy,\cite{morga2023intention} they are implicitly imposing a treatment policy meta-analytical estimand.\cite{vuong2024estimands}

Admittedly, selecting an estimand based on the available data counters the ICH E9 (R1) ``estimand thinking process'', where the estimand is selected first, and data collection is performed to align with the estimand.\cite{mutze2025principles,fletcher2022marking} This is a by-product of the evidence synthesis context, where meta-analyses are often performed based on secondary data of previously conducted studies. A notable exception is the planned ``integrated analysis'' of pivotal RCTs, typically pre-specified prior to data collection, for which the cross-trial alignment of estimands and the definition of a target meta-analytical estimand might be less challenging.\cite{hedman2024estimand}

\section{Case study: semaglutide versus dulaglutide for T2D}\label{sec4}

To illustrate the use of the proposed framework in Section \ref{sec3}, we apply it to a network meta-analysis comparing the efficacy of semaglutide versus dulaglutide in patients with Type 2 diabetes (T2D), based on prior work by Lingvay et al, who performed an exhaustive network meta-analysis, including meta-regression and several scenario analyses.\cite{lingvay2022indirect} This network meta-analysis was selected because T2D is the therapeutic area including some of the first RCTs with information on estimands in trial publications, protocols and statistical analysis plans. 

% This network meta-analysis was selected because T2D is the therapeutic area that motivated many of the regulatory discussions which influenced ICH E9 (R1),

\subsection{Background and evidence base}\label{subsec41}

Long-acting glucagon-like peptide-1 receptor agonists (GLP-1 RAs) such as semaglutide and dulaglutide are established treatment options for patients with T2D. Current treatment guidelines recommend the use of GLP-1 RAs as an effective treatment option for patients with T2D not adequately controlled on metformin alone.\cite{buse20202019} Semaglutide and dulaglutide, both available for once-weekly administration via subcutaneous injection, can provide improved glycemic control for T2D patients.\cite{leiter2017efficacy} Semaglutide and dulaglutide were initially approved for lower-dose maintenance use: 0.5 mg and 1.0 mg for semaglutide, and 0.75 mg and 1.5 mg for dulaglutide. A head-to-head comparison of both treatments, at these doses, was performed in the open-label, multinational, Phase 3b RCT, SUSTAIN 7 (ClinicalTrials.gov registration number NCT02648204), with semaglutide demonstrating superior glycemic control than dulaglutide at week 40 and a similar safety profile.\cite{pratley2018semaglutide} 

The results of SUSTAIN 7 and other trials suggest that higher GLP-1 RA doses can provide increased efficacy, with treatment intensification often required for the maintenance of glycemic control.\cite{fonseca2009defining} Consequently, higher maintenance doses have been investigated, with semaglutide 2.0 mg approved by regulators based on the results of the double-blind, multinational, Phase 3b RCT, SUSTAIN FORTE (NCT03989232);\cite{frias2021efficacySEMA} and dulaglutide 3.0 mg and 4.5 mg approved by regulators based on the results of the double-blind, multinational, Phase 3 RCT, AWARD-11 (NCT03495102).\cite{frias2021efficacy} The absence of head-to-head data between the higher doses of semaglutide and dulaglutide motivated Lingvay et al to perform an indirect treatment comparison between semaglutide 2.0 mg, dulaglutide 3.0 mg and dulaglutide 4.5 mg, based on the results of SUSTAIN 7, SUSTAIN FORTE and AWARD-11.\cite{lingvay2022indirect} Understanding how the higher-dose GLP-1 RAs compare is of interest to clinicians and payers. 

Table \ref{PICO-diabetes} displays the PICO determining the scope of the research question. In this particular case, and contrary to Section \ref{subsec31}, dosing regimens are included for the intervention and comparators because decision-makers are specifically interested in the impact of dosing regimen on clinical outcomes. Two additional comparators, semaglutide 1.0 mg and dulaglutide 1.5 mg, have been included to connect or ``anchor'' the evidence network, as recommended by evidence synthesis methodological guidance.\cite{evisynth} For convenience, we have defined semaglutide 2.0 mg as the intervention, but dulaglutide 3.0 mg or 4.5 mg could have also been defined as the intervention with semaglutide 2.0 mg as a comparator. 

\begin{figure}[!htb]
\centering
\includegraphics[width=12.2cm]{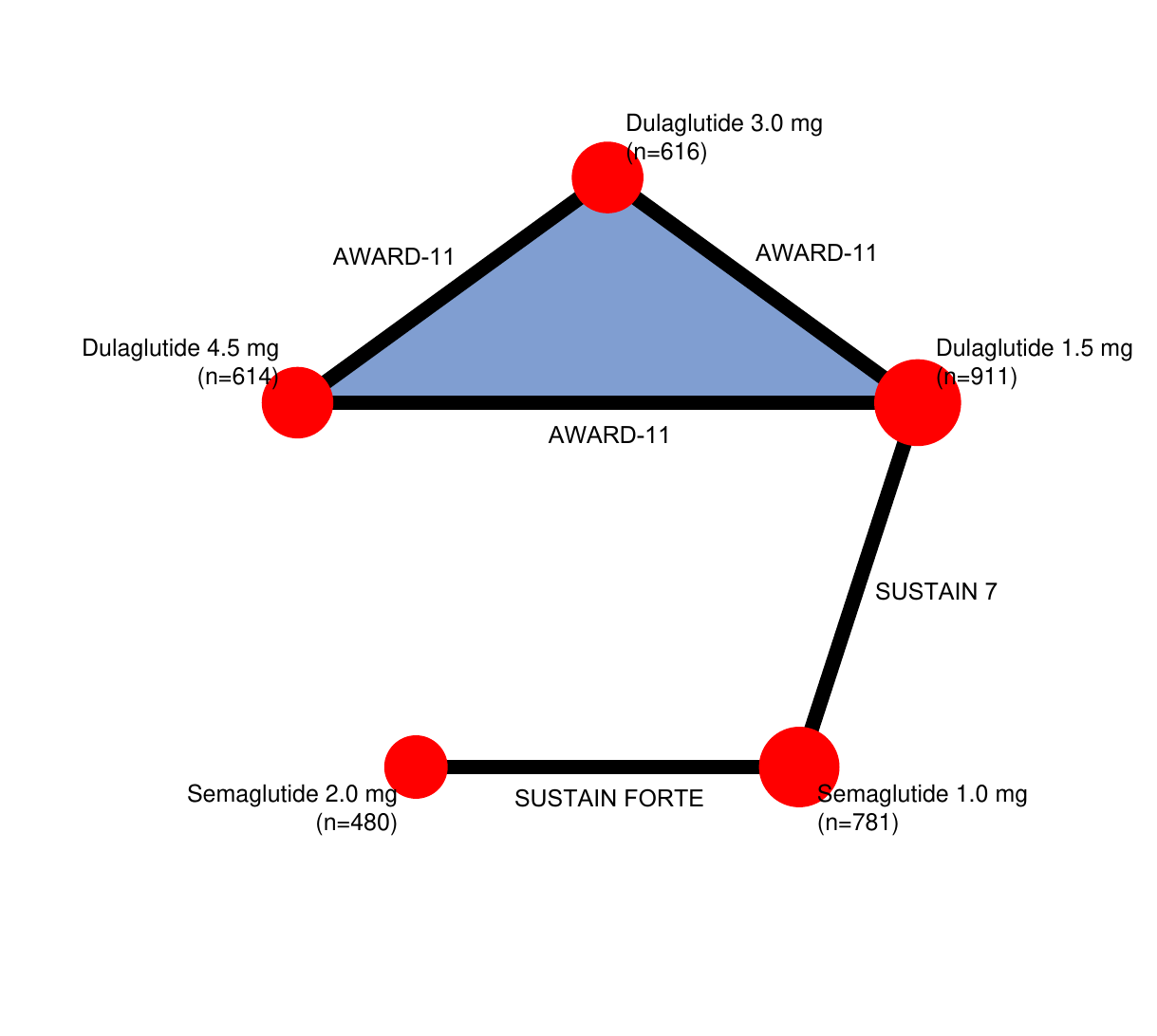}
\caption{Connected evidence network for the treatments in the PICO scope for the case study. The nodes represent different treatments, the edges connecting the nodes indicate the presence of a direct head-to-head comparison in an RCT, and the shaded region highlights a closed loop due to a trial contributing more than two treatment arms (AWARD-11). The sample size $n$ corresponds to the number of subjects randomized to treatment.}
\label{evidence-network}
\end{figure}

Figure \ref{evidence-network} displays the evidence network. SUSTAIN FORTE compares semaglutide 2.0 mg versus semaglutide 1.0 mg; AWARD-11 compares dulaglutide 4.5 mg and 3.0 mg versus dulaglutide 1.5 mg; and SUSTAIN 7 compares semaglutide 1.0 mg versus dulaglutide 1.5mg. The evidence network is connected by including SUSTAIN 7, which anchors the comparisons between the higher-dose GLP-1 RAs via the lower-dose semaglutide 1.0 mg and dulaglutide 1.5 mg.

\begin{table}[!htb]
\centering
\begin{tabular}{|m{10em}||m{30em}|} 
  \hline
  Population & Subjects with Type 2 diabetes on a background treatment of metformin \\   
  \hline
  Intervention & Semaglutide 2.0 mg QW  \\   
  \hline
  Comparators &  
  Dulaglutide 4.5 mg QW \newline
  Dulaglutide 3.0 mg QW \newline 
  \textit{Dulaglutide 1.5 mg QW} \newline
  \textit{Semaglutide 1.0 mg QW} 
  \\ 
  \hline
  Outcomes & Change from baseline in HbA1c \newline
  Change from baseline in body weight \\  
  \hline
\end{tabular}
\caption{PICO defining the scope and broad clinical research question for the case study. The comparators in italics have been included to connect the network of evidence. HbA1c: Hemoglobin A1C (glycated hemoglobin); QW: once a week.}
\label{PICO-diabetes}
\end{table}

\subsection{Estimands at the trial level}\label{subsec42}

We now adopt the estimand framework to systematically identify misalignments between the treatment effects targeted by different RCTs, and attempt to mitigate any quantitative heterogeneity that may threaten the exchangeability of treatment effects across trials and the feasibility of a network meta-analysis. While this exercise may appear reminiscent of general effect modification assessment in risk-of-bias tools for evidence synthesis,\cite{lunny2025risk} it enhances these approaches by accounting for estimand attributes, in particular for intercurrent event strategies. 

\clearpage

Table \ref{trial-level.estimands} displays the trial-level estimands for the RCTs in the evidence base, derived from their publications, protocols and statistical analysis plans.\cite{pratley2018semaglutide, frias2021efficacySEMA, frias2021efficacy, sustain7protocol,sustain7sap, sustainforteprot, sustainfortesap, award11protocol, award11sap} On an attribute-by-attribute basis: 

\begin{itemize}
\item \textbf{Population:} The RCT populations are similar. All trial populations include T2D patients on a background treatment of (i.e., inadequately controlled with) metformin $\geq 1500$ mg/d (or a maximal tolerated dose) $\geq 90$ days or 3 months, with or without sulphonylurea treatment (not explicitly specified for AWARD-11  and SUSTAIN 7). Nevertheless, there are some minor differences in selection criteria, not apparent from the estimands' population attributes. Namely, the baseline HbA1c inclusion criteria are 7.0-10.5\%, 8.0-10.0\% and 7.5\%-11.0\% for SUSTAIN 7, SUSTAIN FORTE and AWARD-11, respectively; AWARD-11 explicitly requires a baseline body mass index $\geq 25$ kg/m\textsuperscript{2}; and SUSTAIN FORTE and AWARD-11 explicitly require T2D duration of at least 6 months. In any case, Lingvay et al (and ClinicalTrials.gov) report generally similar summary statistics for the baseline characteristics of enrolled patients in all RCTs.\cite{lingvay2022indirect}
\item \textbf{Treatment}: The dosing regimen and mode of administration of the common comparators, semaglutide 1.0 mg and dulaglutide 1.5 mg, are identical -- subcutaneous QW -- across the trials in which they are investigated. All treatments in all trials are taken alongside a background of metformin (as specified for the population attribute). 
\item \textbf{Variable (endpoint)}: Change from baseline in HbA1c in percentage points is a primary endpoint for all trials, and change from baseline in body weight in kilograms is a key or confirmatory secondary endpoint for all trials. However, the endpoints are measured at week 40 for the SUSTAIN RCTs, and at week 36 for AWARD-11. This divergence is likely more problematic for body weight than for HbA1c, which is measured over a 12-week time frame. 

\item \textbf{Population-level summary measure}: For all trials, a mean difference between treatment conditions is reported. 
\item \textbf{Intercurrent event strategies}: Overall, two intercurrent events are identified across all trials: premature treatment discontinuation and initiation of anti-diabetic rescue medication. For each of the key primary and secondary trial objectives, all RCTs report estimates of two different estimands, one where both intercurrent events were handled by the treatment policy strategy and another where both were handled by a hypothetical strategy. The similarities and differences between these strategies might not be immediately apparent to readers of the clinical trial journal articles,\cite{pratley2018semaglutide, frias2021efficacySEMA, frias2021efficacy} particularly to those unfamiliar with strategies for intercurrent events, as different terminology has been adopted across trials. We describe the intricacies of the intercurrent event strategies below. 
\end{itemize}

\subsubsection{Intercurrent event strategies}\label{subsec421}

Any similarities and differences between intercurrent event strategies must be made explicit to avoid mixing ``apples and oranges'' in network meta-analyses. The statistical analysis plan of SUSTAIN 7 features a ``de-jure'' and a ``de-facto'' estimand; \cite{sustain7sap} that of SUSTAIN FORTE includes a ``hypothetical'' (denoted ``trial product'' in the corresponding publication)\cite{frias2021efficacySEMA} and a ``treatment policy'' estimand;\cite{sustainfortesap} and that of AWARD-11 describes an ``efficacy'' and a ``treatment regimen'' estimand.\cite{award11sap} The de-jure, hypothetical and efficacy estimands were the primary estimands, except in the USA, where the Food and Drug Administration (FDA) had a preference for the de-facto, treatment policy and treatment regimen estimands. 

The differences in estimand names across trials are due to changes in terminology over time:``de-jure'' and ``de-facto'' estimands refer to ``efficacy'' (effect if treatments are taken as directed in the protocol) and ``effectiveness'' (effect of treatment as actually taken), respectively, but now these terms are somewhat outdated. From the full estimand descriptions in the trial statistical analysis plans,\cite{sustain7sap, sustainfortesap, award11sap} we find that all RCTs contain: (1) a primary estimand, which characterizes the treatment effect if patients do not discontinue treatment prematurely or initiate anti-diabetic rescue medication, i.e., a hypothetical strategy is adopted for these intercurrent events; and (2) a supplementary estimand, requested by the FDA, characterizing the treatment effect regardless of premature treatment discontinuation or initiation of anti-diabetic rescue medication, i.e., a treatment policy strategy is adopted. 

While the criteria for initiation of anti-diabetic rescue medication may have differed across trials, the rescue medication itself was similar. In the SUSTAIN RCTs, it was selected in accordance with the American Diabetes Association and the European Association for the Study of Diabetes guidelines,\cite{davies2018management} but excluding other GLP-1 RAs, dipeptidyl peptidase-4 (DPP-4) inhibitors and amylin analogues.\cite{sustain7protocol, sustainforteprot} In AWARD-11, the rescue therapy was determined by the investigator, but GLP-1 RAs and DPP-4 inhibitors were also excluded.\cite{award11protocol}

It is worth noting that the intercurrent event strategies, while very similar, are not exactly identical across trials. The estimands of SUSTAIN FORTE adopt a treatment policy strategy for change in treatment dose. Conversely, SUSTAIN 7 and AWARD-11 do not explicitly consider change in treatment dose as an intercurrent event. While these two approaches may appear consistent, SUSTAIN FORTE listing change in treatment dose as an intercurrent event, addressed via treatment policy, means that every effort has been made to collect all outcome data after the occurrence of the intercurrent event.\cite{bell2025estimation} Because such intercurrent event was not defined for SUSTAIN-7 and AWARD-11, it is not possible to know how many subjects changed the treatment dose in these trials. The frequency of intercurrent events can change over settings and over time, and an overview of the intercurrent event distribution in the trials would be necessary to understand the impact of ignoring change in treatment dose.\cite{lanius2025realizing} 

As illustrated in Table \ref{estimation-RCT}, different estimation methods are required for different intercurrent event strategies. Selected estimators of the hypothetical strategy set post-intercurrent event outcomes as missing, then impute them to recreate what the outcomes would have been had the intercurrent event not occurred. The imputation is either implicit (mixed model for repeated measures in SUSTAIN 7 and AWARD-11) or explicit (multiple imputation followed by analysis of covariance in SUSTAIN FORTE), relying on a strong ``missing-at-random'' (MAR) assumption, where missing outcomes depend only upon observed and modeled covariates as well as previous outcomes, within randomized treatment group.

Conversely, estimators of the treatment policy strategy use all observed outcomes, and require collecting outcome data following the occurrence of intercurrent events. The estimation approach for the treatment policy strategy in all RCTs is retrieved dropout, which distinguishes between outcomes occurring before and after the intercurrent event, making use of observed post-intercurrent event data to impute missing post-intercurrent event outcomes.\cite{bell2025estimation} This allows for a more flexible ``extended MAR'' assumption, as described by Bell et al,\cite{bell2025estimation} which assumes MAR within group defined by randomized treatment and treatment status (intercurrent event occurrence and timing).\cite{bell2025estimation} All RCTs implemented retrieved dropout by multiple imputation, conditional on whether outcomes occur pre- or post-intercurrent event, followed by analysis of covariance. 

\clearpage

\begin{landscape}

\begin{table}
\centering
\begin{tabular}{|m{10em}|m{10em}| m{10em}|m{10em}|m{14em}|} 
  \hline
  Trial & Estimand name & Strategy for initiation of rescue medication and treatment discontinuation & Statistical model & Imputation approach and assumption \\ 
  \hline
  \hline
  \multirow{2}{*}{AWARD-11\cite{frias2021efficacy,   award11protocol, award11sap}} & Efficacy & Hypothetical & MMRM & MMRM: MAR within randomized treatment group \\\cline{2-5} 
  & Treatment regimen  & Treatment policy
   & Retrieved dropout (ANCOVA) & MI: (extended) MAR within group defined by randomized treatment and treatment status at week 36 \\ 
  \hline
  \multirow{2}{*}{SUSTAIN 7\cite{pratley2018semaglutide,sustain7protocol, sustain7sap}} & De-jure & Hypothetical & MMRM & MMRM: MAR within randomized treatment group \\\cline{2-5}
  & De-facto & Treatment policy & Retrieved dropout (ANCOVA) & MI: (extended) MAR within group defined by randomized treatment and treatment status at week 40 \\   
  \hline
  \multirow{2}{*}{\parbox{1.5cm}{SUSTAIN FORTE\cite{frias2021efficacySEMA,sustainforteprot, sustainfortesap}}} & Hypothetical (trial product) & Hypothetical & ANCOVA & MI: MAR within randomized treatment group \\\cline{2-5}
  & Treatment policy & Treatment policy & Retrieved dropout (ANCOVA) & MI: (extended) MAR within group defined by randomized treatment and treatment status at week 40 \\ 
  \hline
\end{tabular}
\caption{Statistical estimation procedures for the trials included in the evidence base of the case study. ANCOVA: analysis of covariance; MAR: missing-at-random; MI: multiple imputation; MMRM: mixed model for repeated measures.}
\label{estimation-RCT}
\end{table}

\clearpage

\begin{table}
\centering
\begin{tabular}{|m{5.5em}||m{8.5em}|m{8.5em}|m{8.5em}|m{8.5em}|m{8.5em}|m{8.5em}|} 
  \hline
  Trial & \multicolumn{2}{c|}{AWARD-11\cite{frias2021efficacy,   award11protocol, award11sap} (NCT03495102)} & \multicolumn{2}{c|}{SUSTAIN 7\cite{pratley2018semaglutide,sustain7protocol, sustain7sap} (NCT02648204)} & \multicolumn{2}{c|}{SUSTAIN FORTE\cite{frias2021efficacySEMA,sustainforteprot, sustainfortesap} (NCT03989232)}\\   
  \hline
  Estimand name & Efficacy & Treatment regimen & De-jure & De-facto & Hypothetical (trial product) & Treatment policy \\ 
  \hline
  Population & \multicolumn{2}{p{17em}|}{Subjects with Type 2 diabetes inadequately controlled with metformin} & \multicolumn{2}{p{17em}|}{Subjects with Type 2 diabetes on a background treatment with metformin} & \multicolumn{2}{p{17em}|}{Subjects with Type 2 diabetes on a background treatment of metformin with or without sulphonylurea treatment} \\
  \hline
  Treatments (on top of metformin) & \multicolumn{2}{p{17em}|}{Subcutaneous dulaglutide 4.5 mg QW \newline
  Subcutaneous dulaglutide 3.0 mg QW \newline 
  Subcutaneous dulaglutide 1.5 mg QW} & \multicolumn{2}{p{17em}|}{Subcutaneous semaglutide 1.0 mg QW \newline Subcutaneous semaglutide 0.5 mg QW 
  \newline
  Subcutaneous dulaglutide 1.5 mg QW \newline 
  Subcutaneous dulaglutide 0.75 mg QW}
  & \multicolumn{2}{p{17em}|}{Subcutaneous semaglutide 2.0 mg QW \newline Subcutaneous semaglutide 1.0 mg QW}
  \\  
  \hline
  Variables (endpoints) & 
\multicolumn{2}{p{17em}|}{Primary: 
change from baseline to week 36 in HbA1c (\%-points) \newline
  Key secondary: change from baseline to week 36 in body weight (kg)} & \multicolumn{2}{p{17em}|}{Primary: change from baseline to week 40 in HbA1c (\%-points) \newline
  Confirmatory secondary: change from baseline to week 40 in body weight (kg)} & \multicolumn{2}{p{17em}|}{Primary: change from baseline to week 40 in HbA1c (\%-points) \newline
  Confirmatory secondary: change from baseline to week 40 in body weight (kg)} \\
  \hline
  Population-level summary measure & \multicolumn{2}{p{17em}|}{Mean difference between treatment conditions} &   \multicolumn{2}{p{17em}|}{Mean difference between treatment conditions} &   \multicolumn{2}{p{17em}|}{Mean difference between treatment conditions}\\
  \hline
  Intercurrent event strategies & \textbf{Hypothetical}: \newline  \textbullet\hspace{0.2cm} Initiation of anti-diabetic rescue medication 
  \newline \textbullet\hspace{0.2cm} Premature treatment discontinuation  
  & 
  \textbf{Treatment policy}: \newline  \textbullet\hspace{0.2cm} Initiation of anti-diabetic rescue medication 
  \newline \textbullet\hspace{0.2cm} Premature treatment discontinuation  
  & 
\textbf{Hypothetical}: \newline  \textbullet\hspace{0.2cm} Initiation of anti-diabetic rescue medication 
  \newline \textbullet\hspace{0.2cm} Premature treatment discontinuation  
  & 
  \textbf{Treatment policy}: \newline  \textbullet\hspace{0.2cm} Initiation of anti-diabetic rescue medication 
  \newline \textbullet\hspace{0.2cm} Premature treatment discontinuation
  & 
\textbf{Hypothetical}: \newline  \textbullet\hspace{0.2cm} Initiation of anti-diabetic rescue medication 
  \newline \textbullet\hspace{0.2cm} Premature treatment discontinuation  \newline
  \textbf{Treatment policy}:\newline
\textbullet \hspace{0.2cm}Change in treatment dose &
\textbf{Treatment policy}: \newline 
\textbullet\hspace{0.2cm}Initiation of anti-diabetic rescue medication 
  \newline \textbullet\hspace{0.2cm} Premature treatment discontinuation  \newline
\textbullet \hspace{0.2cm}Change in treatment dose
  \\
  \hline 
\end{tabular}
\caption{Trial-level estimands, according to the ICH E9 (R1) estimand framework, of the RCTs in the evidence base for the case study.}
\label{trial-level.estimands}
\end{table}

\clearpage

\end{landscape}

\subsection{Estimands at the meta-analytical level}\label{subsec43}

We now attempt to define estimands at the meta-analytical level, aiming to enhance the relevance of pooled treatment effect estimates with respect to a specific clinical research question, and to maximize their applicability to healthcare decision-making contexts. There is some cross-trial heterogeneity between the estimand attributes in the case study; particularly, in the time at which endpoints are measured. Nevertheless, other sources of heterogeneity have been mitigated, e.g., by identifying consistent intercurrent event strategies across studies.

\begin{figure}[htb]
\centering\includegraphics[width=0.75\linewidth]{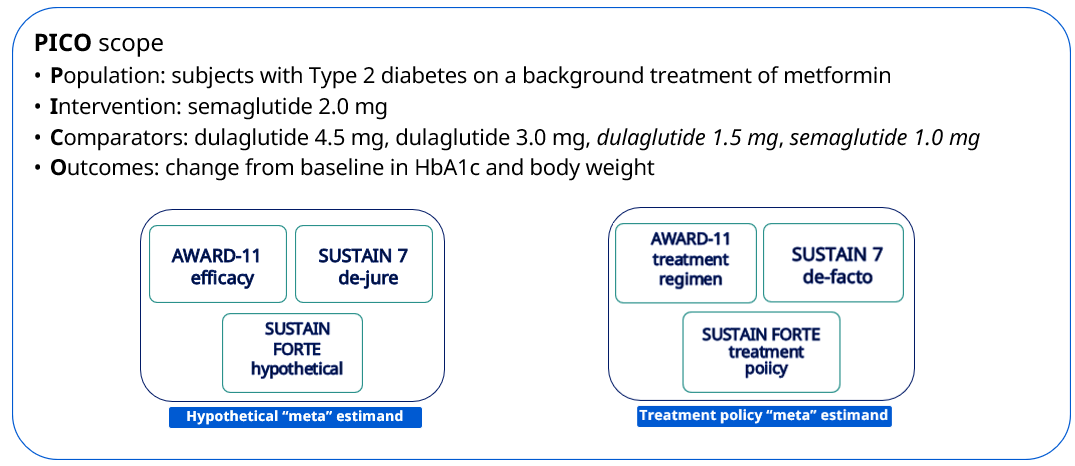}
  \caption{The evidence base, defined by the PICO scope, can be narrowed using each meta-analytical estimand to perform more targeted network meta-analyses that are relevant to specific clinical research questions.}
  \label{PICO_restriction}
\end{figure}

We construct two target meta-analytical estimands (Table \ref{PICO-diabetes-2}): one which adopts a hypothetical strategy for premature treatment discontinuation or initiation of anti-diabetic rescue medication (the hypothetical meta-analytical estimand), consistent with the de-jure, hypothetical (trial product) and efficacy estimands from SUSTAIN 7, SUSTAIN FORTE and AWARD-11, respectively; and another which adopts a treatment policy strategy for premature treatment discontinuation or initiation of anti-diabetic rescue medication (the treatment policy meta-analytical estimand), consistent with the de-facto, treatment policy and treatment regimen estimands from SUSTAIN 7, SUSTAIN FORTE and AWARD-11, respectively. Figure \ref{PICO_restriction} illustrates the restriction of the original PICO scope using each target meta-analytical estimand. 

\begin{table}[!htb]
\centering
\begin{tabular}{|m{14em}||m{16em}|m{16em}|} 
  \hline
  Estimand name & Hypothetical & Treatment policy \\   
  \hline
  Population & \multicolumn{2}{p{32em}|}{Subjects with Type 2 diabetes on a background treatment of (i.e., inadequately controlled with) metformin} \\
  \hline
  Treatments &  \multicolumn{2}{p{32em}|}{
  Semaglutide 2.0 mg QW \newline
  Dulaglutide 4.5 mg QW \newline
  Dulaglutide 3.0 mg QW \newline 
  \textit{Dulaglutide 1.5 mg QW} \newline
  \textit{Semaglutide 1.0 mg QW} }
  \\ 
  \hline
  Variables (endpoints) & \multicolumn{2}{p{32em}|}{Change from baseline to week 36 or week 40 in HbA1c (\%-points)
  \newline
  Change from baseline to week 36 or week 40 in body weight (kg)  
  }
  \\  
  \hline
  Population-level summary measure & \multicolumn{2}{p{30em}|}{Mean difference between treatment conditions
  }
  \\  
  \hline
   Intercurrent event strategies & 
\textbf{Hypothetical} strategy for initiation of anti-diabetic rescue medication or premature treatment discontinuation  
  & \textbf{Treatment policy} strategy for initiation of anti-diabetic rescue medication or premature treatment discontinuation \\  
  \hline
\end{tabular}
\caption{Target meta-analytical estimands for the case study. The treatments in italics were not of primary interest and were included to connect the evidence network.}
\label{PICO-diabetes-2}
\end{table}

\clearpage

\subsubsection{Example network meta-analyses}\label{subsec431}

We conduct two network meta-analyses, exploring the impact of targeting different meta-analytical estimands or intercurrent event strategies. One of the network meta-analyses adopts a hypothetical strategy for premature treatment discontinuation or initiation of anti-diabetic rescue medication. The other adopts a treatment policy strategy for the corresponding intercurrent events. T2D is a therapeutic area where confirmatory Phase 3 RCTs typically report results for both hypothetical and treatment policy estimands. This allows us to conduct the two network meta-analyses based on published aggregate-level results in the clinical trial publications,\cite{pratley2018semaglutide, frias2021efficacySEMA, frias2021efficacy} without necessarily requiring access to subject-level data.

\begin{figure}[!htb] 
    \centering
    \subfloat[\centering Hypothetical meta-analytical estimand]{{\includegraphics[width=6.45cm]{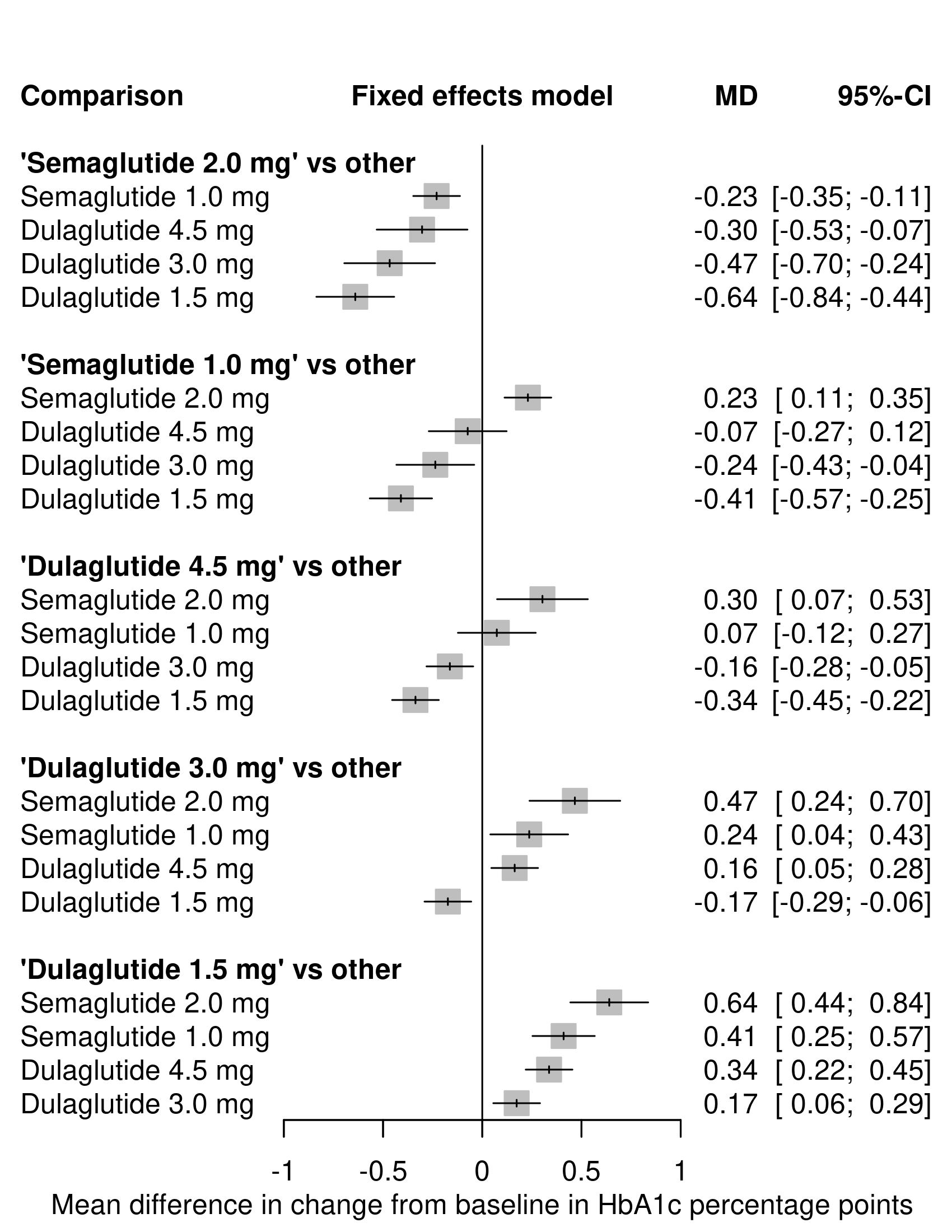} }}%
    \qquad
    \subfloat[\centering Treatment policy meta-analytical estimand]{{\includegraphics[width=6.45cm]{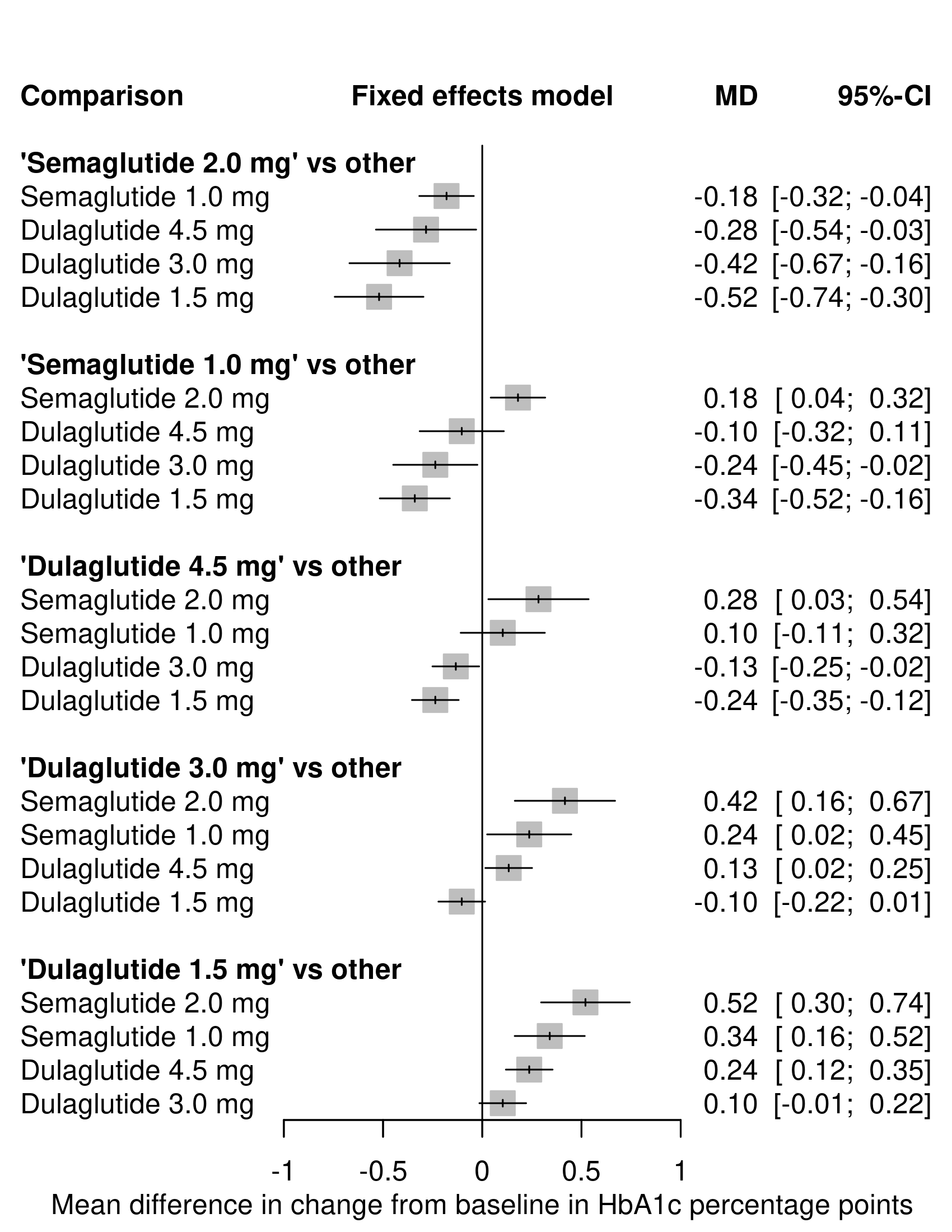} }}%
    \caption{Results of the network-meta analyses for change from baseline in HbA1c (\%-points). MD: mean difference; CI: confidence interval.}%
    \label{hba1c results}%
\end{figure}

\begin{figure}[!htb]%
    \centering
    \subfloat[\centering Hypothetical meta-analytical estimand]{{\includegraphics[width=6.45cm]{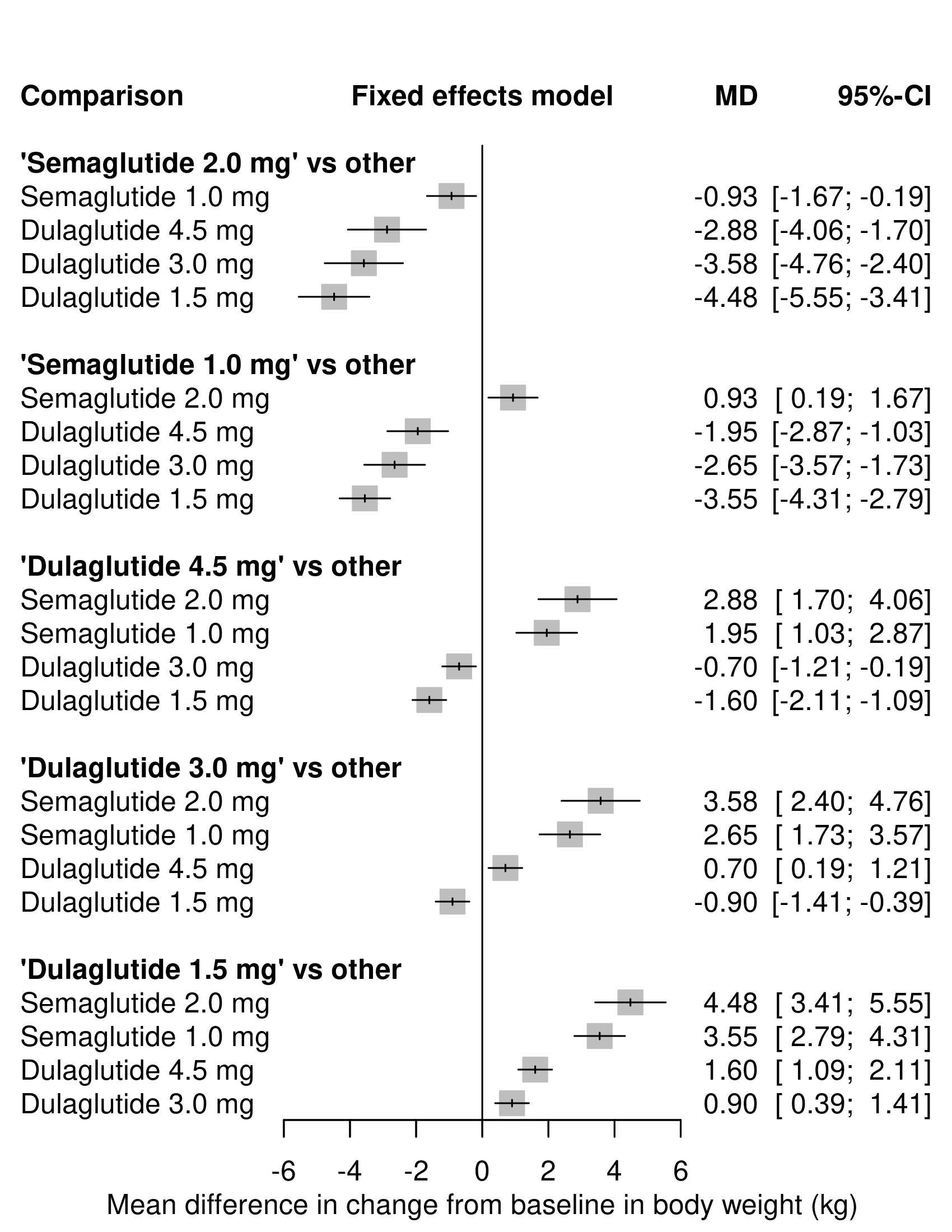} }}%
    \qquad
    \subfloat[\centering Treatment policy meta-analytical estimand]{{\includegraphics[width=6.45cm]{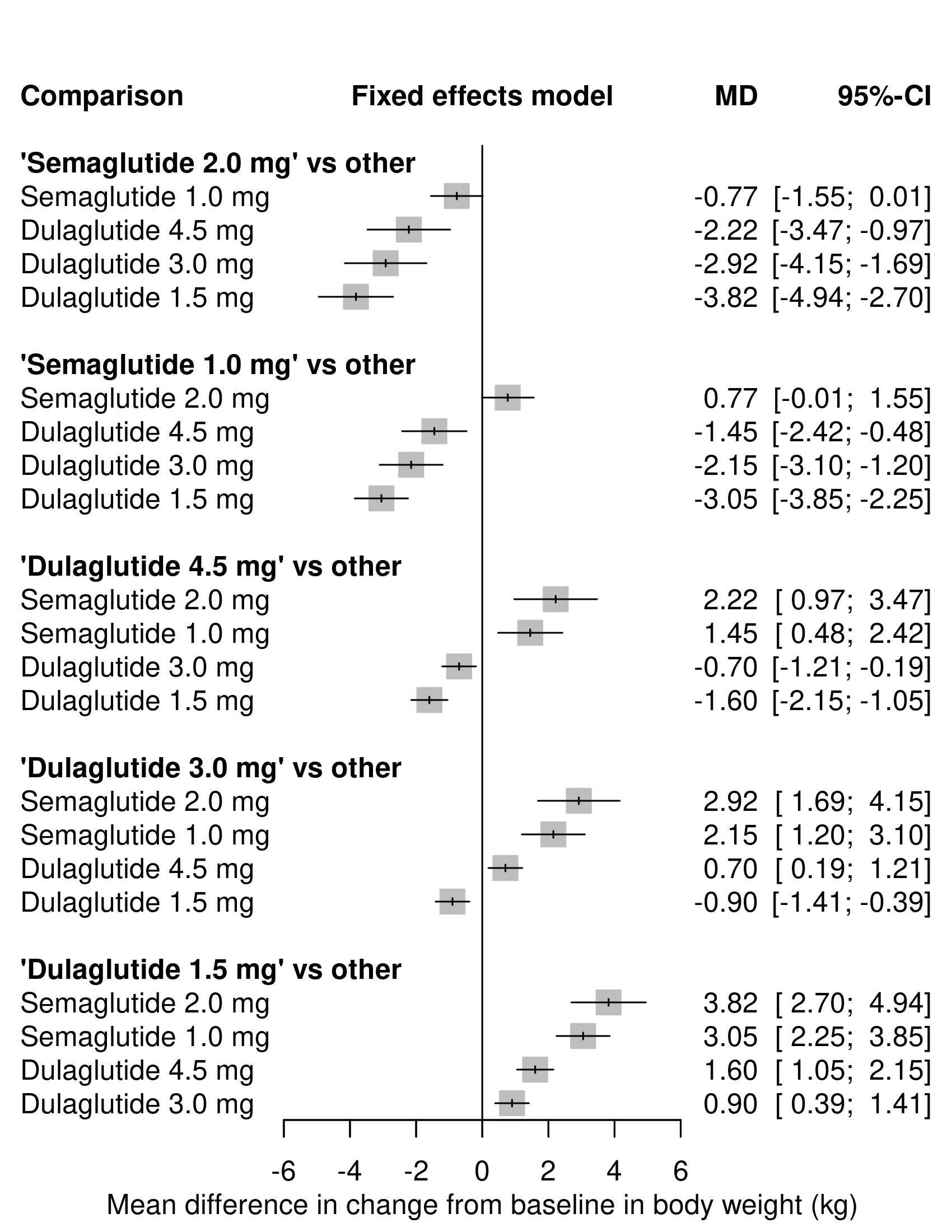} }}%
    \caption{Results of the network-meta analyses for change from baseline in body weight (kg). MD: mean difference; CI: confidence interval.}%
    \label{body weight results}%
\end{figure}

The network meta-analyses are performed based on trial-level mean differences and standard errors, using fixed effects models with inverse variance weighting,\cite{borenstein2010basic} assuming independence between trials and accounting for correlated treatment effects within multi-arm trials,\cite{rucker2012network} as implemented in the R package netmeta.\cite{balduzzi2023netmeta} Where standard errors were not reported for treatment effects (treatment regimen estimates for AWARD-11), they were approximated based on reported confidence intervals for mean change from baseline for each treatment arm.\cite{higgins2019cochrane} The results of the network meta-analyses for change from baseline in HbA1c, targeting the hypothetical and treatment policy estimands at the ``meta'' level, are in Figure \ref{hba1c results}. The corresponding results for change from baseline in body weight are in Figure \ref{body weight results}. 

For change from baseline in HbA1c, estimated mean differences for semaglutide 2.0 mg versus dulaglutide 3.0 mg and 4.5 mg are -0.47 (95\% confidence interval [CI] -0.70 to -0.24) and -0.30 (95\% CI -0.53 to -0.07) percentage points, respectively, for the hypothetical strategy; and -0.42 (95\% CI -0.67 to -0.16) and -0.28 (95\% CI -0.54 to -0.03) percentage points, respectively, for the treatment policy strategy. For change from baseline in body weight, estimated mean differences for semaglutide 2.0 mg versus dulaglutide 3.0 mg and 4.5 mg are -3.58 (95\% CI -4.76 to -2.40) kilograms and -2.88 (95\% CI -4.06 to -1.70) kilograms, respectively, for the hypothetical strategy; and -2.92 (95\% CI -4.15 to -1.69) and -2.22 (95\% CI -3.47 to -0.97) kilograms, respectively, for the treatment policy strategy. 

At the RCT level, and as indicated by the results of the individual trials,\cite{pratley2018semaglutide, frias2021efficacySEMA, frias2021efficacy} treatment effects under a treatment policy strategy are generally more attenuated than those under a hypothetical strategy. Because the effects of intercurrent events tend to be common across arms, subjects experiencing an intercurrent event tend to become more similar.\cite{ratitch2020choosing} In diabetes GLP-1 RA clinical trials, including measurements that are collected after treatment discontinuation makes the actual treatment received, on average, more alike between the experimental and the control arms.\cite{aroda2019pioneer} The initiation of anti-diabetic rescue medication is, typically, only allowed if patients do not maintain acceptable glycemic control. Because patients are less likely to maintain adequate glycemic control under the less effective treatment, the initiation of rescue medication is expected to occur more commonly among participants receiving such treatment.\cite{aroda2019pioneer} Hence, including measurements collected after the intercurrent event as part of a treatment policy strategy generally leads to a smaller difference between the treatment arms of a trial.\cite{aroda2019pioneer} 

The magnitude of the hypothetical and treatment policy pooled estimates is similar for change from baseline in HbA1c at the meta-analytical level (Figure \ref{hba1c results}), with the treatment policy estimate slightly closer to the null. For change from baseline in body weight, the pull towards the null by the treatment policy strategy is more noticeable (Figure
5), more in line with what would be expected in an individual RCT. 

% Trial-level HbA1c treatment effects appear to have been similarly attenuated across studies, with dilutions averaging out across the evidence network. Conversely, for body weight, the pull towards the null is stronger in some trials (SUSTAIN 7) than in others (SUSTAIN FORTE and AWARD-11). 

\subsubsection{Applicability of meta-analytical estimands}\label{subsec432}

HTA concerns the evaluation of relative effectiveness (added treatment benefit in conditions encountered in ``real-world'' clinical practice), as opposed to clinical efficacy (added treatment benefit under ideal, controlled, circumstances).\cite{nordon2016efficacy} The choice of target meta-analytical estimand, and its intercurrent event strategy, should account for external validity or applicability to the decision-making context; in the HTA case, whether treatment effects can be reasonably applied to patients in routine clinical practice.  

If treatment discontinuation and the initiation of anti-diabetic rescue medication are occurring in real-world clinical practice, the treatment policy meta-analytical estimand is likely of greater interest to HTA decision-makers. Nevertheless, decision-makers must consider that treatment discontinuation rates and rescue medication can vary over time, settings and jurisdictions. This may threaten the applicability of treatment policy results, which depends on each intercurrent event and its distribution being stable. There are also concerns about informative missingness if post-intercurrent event outcomes are not collected. For the RCTs in our case study, treatment discontinuation rates at week 36 or 40 were 7\%-17\% for the treatments in the scope, which is roughly consistent with the 20\% discontinuation rate observed after one year for T2D populations using GLP-1 RAs in European real-world settings,\cite{palanca2023real, lassen2024adherence} but considerably lower than the 50\% observed after one year in the United States and Israel.\cite{weiss2020real, kassem2024efficacy}  

\clearpage

\clearpage

The hypothetical meta-analytical estimand targets the treatment effect if patients adhere to treatment as intended, without treatment discontinuation or rescue medication.\cite{aroda2019pioneer} It might still be of interest to HTA decision-makers if the reasons for intercurrent events are not related to clinical effectiveness (including safety), or if there is a clear manner or incentive to prevent the occurrence of intercurrent events in future practice. This is challenging for treatment discontinuation here. While some T2D patients in routine clinical practice discontinue subcutaneous GLP-1 RAs for cost- or preference-related reasons, e.g., preferring oral medication over injections, many discontinue treatment due to inadequate glycemic control, nausea or gastrointestinal events.\cite{sikirica2017reasons} 

Ultimately, T2D patients do prematurely discontinue GLP-1 RAs in routine clinical practice, at rates generally higher than those observed in RCTs, and for efficacy or safety reasons that cannot be prevented trivially.\cite{lim2025treatment} As such, an estimand that envisages a hypothetical scenario where patients do not discontinue treatment likely has limited applicability to HTA decision-making. Such estimand would have been targeted at the meta-analytical level by analysts na\"ively pooling the primary estimands of the RCTs. This is something that probably happens in current meta-analytical practice, where different intercurrent event strategies are typically ignored.\cite{metcalfe2025treatment} 

Interestingly, an European Medicines Agency guideline for the use of estimands in diabetes trials\cite{EMAdiabetesguideline}  recommends a treatment policy strategy for premature treatment discontinuation and a hypothetical strategy for initiation of anti-diabetic rescue medication.\footnote{Another suggestion in the guideline is the use of a composite strategy to account for the two intercurrent events in the variable/endpoint definition, but as part of a supplementary estimand.\cite{aroda2019pioneer, EMAdiabetesguideline}} Namely, the target estimand is the treatment effect if anti-diabetic rescue medication is not initiated, regardless of premature treatment discontinuation. Such estimand is relevant because patients under the standard-of-care in routine clinical practice may discontinue treatment, and while they may initiate anti-diabetic rescue therapy, this is often medication within the same drug class (not permitted in comparative GLP-1 RA trials).\cite{lim2025treatment} In our case study, targeting the ``hybrid'' estimand, both at the trial and at the meta-analytical level, would have required full access to subject-level data because the individual RCTs did not plan or report results for the hybrid intercurrent event strategy. We shall discuss the benefits brought by subject-level data availability in Section \ref{sec5}. 

\section{Discussion}\label{sec5}

\subsection{Subject-level data availability}

Estimands can be helpful to communicate limitations of the evidence base where subject-level data are unavailable. Our case study is in a therapeutic area, T2D, which stimulated the regulatory discussions influencing ICH E9 (R1), and where recent RCT publications typically report both a treatment policy and a hypothetical estimand for premature treatment discontinuation or initiation of anti-diabetic rescue medication. One is unlikely to encounter such standardized reporting and consistency in estimands across many other therapeutic areas, but this is likely to change over time, as we expect most disease-specific regulatory guidance to be updated with recommended estimands. The harmonization of estimands within therapeutic areas has been identified as a priority for future regulatory-HTA body collaborations and should allow for more reliable evidence syntheses.\cite{regulatory-HTAb} 

In any case, it may be challenging in many cases to align all estimand attributes -- particularly, intercurrent event strategies -- without full access to subject-level data. For instance, Table \ref{estimand-cabo} in Section \ref{subsec31} highlighted the primary estimands for a thyroid cancer RCT, which included eight and nine different intercurrent events for each of the two objectives. Aligning on every single trial-level intercurrent event strategy for this indication, especially under subject-level data limitations, will be very challenging. An informed discussion with clinicians will be required to determine which intercurrent events are more relevant, both in terms of frequency and clinical importance. In T2D, premature treatment discontinuation and the initiation of anti-diabetic rescue medication are well-established intercurrent events that should be accounted for, even in aggregate-level meta-analyses. In oncology trials, treatment switching is a major intercurrent event that should be considered.\cite{vuong2024estimands, metcalfe2025treatment} 

Ades et al outline how quantitative heterogeneity for ``outcomes reported in different ways'' can be addressed through multivariate normal random effects meta-analysis.\cite{ades2024twenty} Similar analytical methodologies\cite{bujkiewicz2019bivariate} could be applied where trials report different sets of intercurrent event strategies, e.g., if a relevant intercurrent event strategy is reported in some trials but not in others, multivariate (network) meta-analyses could be used to borrow information from trials in which the relevant estimands are reported, thereby accounting for the totality of evidence.\cite{remiro2024broad} Clinical trial reporting guidelines should continue to be updated to include information on intercurrent events, so there is an incentive to more easily study them in aggregate-level data meta-analyses. In any case, the conduct of analyses based on subject-level data, as opposed to aggregate-level data, should be encouraged.\cite{riley2023using} It will likely become more feasible in the future, as pharmaceutical companies are increasingly allowing access to subject-level data.\cite{modi202310} 

In our case study, the use of subject-level data would have allowed for targeting the hybrid estimand recommended by the European Medicines Agency, described in Section \ref{subsec432}, at the trial and at the meta-analytical level. It would have also allowed us to pursue greater alignment between the timing of outcome assessments in the SUSTAIN trials (week 40) and AWARD-11 (week 36). Nevertheless, even with full access to subject-level data, it can happen that the relevant data are not collected for all trials. For instance, in our case study, no outcome data were collected at week 36 for the SUSTAIN trials and at week 40 for AWARD-11, as there were no visits during those weeks. 

Lingvay et al attempted to derive results for week 36 from the SUSTAIN 7 and SUSTAIN FORTE subject-level data via linear interpolation between week 28 and week 40, which provided results consistent with those of the original analysis.\cite{lingvay2022indirect} Alternatively, one could have used subject-level data from AWARD-11 to derive week 40 results by linear interpolation between the week 36 and week 52 visits, or attempted to model time-course relationships by synthesizing outcomes at multiple visits.\cite{pedder2019modelling, pedder2020performance} Notably, our case study has focused on efficacy endpoints but additional difficulties may arise with safety endpoints due to cross-trial differences in the assessment of adverse events.\cite{unkel2019estimands} 

It is worth noting that adopting common intercurrent event strategies across trials may not always be feasible, even with full access to subject-level data. For instance, alignment with respect to a treatment policy strategy would require all trials to collect outcome data after the occurrence of the intercurrent event. Nevertheless, trials designed with other intercurrent event handling strategies in mind may not collect post-intercurrent event outcome data. A trial prioritizing a composite variable strategy -- which incorporates the occurrence of the intercurrent event into the endpoint definition -- might not have even reported whether it is the primary outcome of interest or the intercurrent event that has occurred for a given subject. 

Consider the design of an RCT for which a treatment policy strategy is not of primary interest to the sponsor, but in a therapeutic area where other trials report treatment policy estimands. We encourage trialists to collect post-intercurrent event outcome data to align with a treatment policy strategy for meta-analysis, with the caveat that the meaning and frequency of intercurrent events might change over time. The value of this alignment likely depends on the decision-making relevance of the meta-analytical treatment policy estimand, the type and frequency of intercurrent events being stable across trials, and the feasibility of aligning other intercurrent event strategies instead (if these are relevant to the research question). We also encourage trialists to report information on the intercurrent event distribution to assess the comparability of treatment policy estimands across trials. 

\subsection{Meta-analysis and meta-regression methodology}

Another limitation of our case study is the slight differences in the trials' inclusion criteria, which imply that the RCT populations are not identical, even though the high-level population attribute in the estimand definitions is virtually the same. It is worth noting that, even if the trials' selection criteria were to be the same, differences in the distribution of baseline characteristics can arise between ``analysis sets'', due to non-random sampling. 

Lingvay et al report summary statistics for the baseline characteristics of enrolled patients in SUSTAIN 7, SUSTAIN FORTE and AWARD-11.\cite{lingvay2022indirect} These are generally similar, but some minor differences exist.\cite{lingvay2022indirect} Based on subgroup analyses of SUSTAIN 7, there is little evidence to suggest that these differences are ``effect-modifying'', with the caveat that the subgroup analyses are post-hoc and underpowered to test for effect modification.\cite{pratley2020impact} Again, an informed discussion with clinicians is generally required to determine the importance of cross-trial differences in the distribution of baseline characteristics. 

Cross-trial differences in the distribution of baseline characteristics can give rise to treatment effect heterogeneity, which compromises the validity of (network) meta-analyses.\cite{phillippo2018methods, phillippo2016nice} Traditional statistical tests used to quantify heterogeneity, such as Cochran’s Q test and the I-squared statistic, do not explain heterogeneity (and are often underpowered). We propose the use of the estimand framework to explicitly describe the different sources that may be inducing such heterogeneity. When unexplained heterogeneity remains, random effects models have traditionally been used to pool treatment effects.\cite{dersimonian2007random, stijnen2010random} However, these could not be applied to our case study, which included only one trial per pairwise treatment comparison. 

Both fixed and random effects approaches have been criticized for not explaining heterogeneity and not explicitly producing directly interpretable estimates in any well-defined target population.\cite{sobel2017causal, dahabreh2020towards} Conventional methods weigh trial-specific estimates according to their precision rather than their relevance to a specific target population, and produce ambiguous estimates with limited external validity.\cite{manski2020toward, hasegawa2017myth} This criticism can be mitigated by narrowly defining the population in the scope,\cite{ades2024twenty, remiro2024broad} such that quantitative heterogeneity between trials is reduced and common effects are more plausible.

If there is no quantitative heterogeneity across trials, both fixed and random effects analyses target a single underlying estimand, common to all studies, for each pairwise treatment comparison. Conversely, under the presence of cross-trial quantitative heterogeneity, e.g., due to differences in patient populations, endpoint timing, etc., there are different trial-specific estimands. Both fixed and random effects models average over the residual heterogeneity, thereby targeting an average effect in the overall population (``super-population'') of included trials compatible with the broader overarching PICO or meta-analytical estimand. 

The difference between fixed and random effects analyses is that the latter can quantify the residual heterogeneity and incorporate it through a predictive distribution.\cite{higgins2009re} Arguably, this implies that treatment effects cannot be interpreted without a description of the full distribution of estimated random effects, not just its center (e.g., average) value.\cite{hasegawa2017myth} Previous research has implied that fixed effects analyses consider a single estimand at the individual trial and meta-analytical levels, for each pairwise treatment comparison.\cite{vuong2024estimands} Nevertheless, this interpretation only holds if there is no heterogeneity across trials, such that the collection of trial-specific estimands is homogeneous, and in which case the interpretation also holds for a random effects analysis. 

Cross-trial discrepancies in populations could have been resolved through innovative meta-regression techniques for covariate adjustment, such as multilevel network meta-regression,\cite{lingvay2022indirect, phillippo2020multilevel, phillippo2020assessing, phillippo2023validating} which would require access to some subject-level data or the use of federated systems.\cite{han2025federated} Meta-regression approaches seem relevant to address the meta-analytical challenges discussed in Section \ref{sec1}, which motivated our proposal for the use of estimands, specifically with respect to the population attribute. They can explain heterogeneity and explicitly mitigate it, by adjusting for cross-trial differences in the distribution of effect-modifying baseline characteristics. We note that -- even if heterogeneity is minimized -- the ``population'' attribute of all trial-specific estimands may not match the target population for decision-making. Meta-regression approaches also hold promise because they can explicitly produce treatment effects in any target population of substantive interest, not necessarily that of one of the trials in the evidence base, thereby enhancing the external validity that is necessary for HTA decision-making. 

\subsection{Population-level summary measure}

Our exposition has focused on one of the estimand attributes that is not considered by PICO, the intercurrent event strategy, and on its alignment across trials. There is another estimand attribute not included in PICO, the population-level summary measure. Conflicts in the marginal and conditional summary measures that are pooled in meta-analyses are a cause for concern, particularly for analyses based on aggregate-level data, as trials may report different summary measures\cite{remiro2025marginal} and na\"ively combining them may produce bias.\cite{remiro2022parametric} For non-collapsible summary measures, such as (log) odds ratios and (log) hazard ratios, marginal and conditional estimands -- and conditional estimands conditioned on different covariate sets -- do not coincide, even in the absence of effect modification by the covariates.\cite{daniel2021making} In the presence of effect modification, this is equally problematic for some collapsible measures, those that are not directly collapsible.\cite{remiro2024transportability} 

Fortunately, the population-level summary measure for all RCTs in our case study is a mean difference from change from baseline, which is directly collapsible.\cite{remiro2024transportability} Table \ref{estimation-RCT} illustrates the trial-level estimation methods that were employed for each RCT. The mixed model for repeated measures (MMRM) estimates the difference in ``least squares'' means between treatment groups, which corresponds to targeting a conditional treatment effect at the covariate means. The modeling approaches based on analysis of covariance (ANCOVA) also target a conditional treatment effect, assumed homogeneous across covariate values and therefore equivalent to the conditional treatment effect at the covariate means.\cite{remiro2025marginal} Again, full access to subject-level data would be beneficial to address cross-trial summary measure incompatibilities, by allowing for the estimation of any desired marginal or conditional summary measure for each of the trials, or by permitting for a subject-level (network) meta-regression. 

Aligning the target summary measure at the meta-analytical level with a relevant research question for decision-making is also crucial. Marginal and conditional summary measures correspond to different decision questions, with discussion over which is more suitable for healthcare decision-making,\cite{Phillippo_Remiro-Azócar_Heath_Baio_Dias_Ades_Welton_2025} and some arguing that marginal summary measures are of greater interest for population-level decision-making.\cite{remiro2025marginal} In the context of our case study, conditional mean differences at the covariate means and marginal mean differences are expected to coincide due to direct collapsibility, as long as conditional effects are constant across covariate values (assumed by the trial-level analytical approaches) or vary linearly with the covariates.\cite{remiro2025marginal, remiro2024transportability} 

\section{Concluding remarks}

Pairwise and network meta-analyses are crucial to inform healthcare policy, but threatened by quantitative heterogeneity between trials and by the poor applicability of pooled estimates to decision-making contexts. We have applied the estimand framework to a case study of semaglutide versus dulaglutide for T2D in order to explicitly identify potential misalignments in the evidence base, mitigate
key sources of quantitative heterogeneity between trials, and to enhance the applicability of meta-analytical results with respect to a specific research question. We performed a network meta-analysis pooling exclusively hypothetical estimates, and a network meta-analysis pooling exclusively treatment policy estimates. This implied the specification of a hypothetical and a treatment policy target estimand at the meta-analytical level, allowing us to be transparent and explicit about the source of heterogeneity -- the intercurrent event strategy -- driving the difference between research questions. 

There were more notable divergences in results between intercurrent event strategies for one endpoint (change from baseline in body weight) than for the other (change from baseline in HbA1c). It is worth noting that we exclusively pooled estimates using the same intercurrent event strategy, such that the treatment policy attenuation of trial-level estimates may have averaged out across the network. In practice, the na\"ive pooling of estimates targeting different intercurrent event strategies, especially where the intercurrent event occurs frequently, may shift efficacy and (cost-)effectiveness results in arbitrary directions while producing estimates of ambiguous interpretability.\cite{vuong2024estimands}

% Numerical differences might be large enough, particularly for body weight, to potentially impact cost-effectiveness estimates such as those used by HTA bodies to make reimbursement decisions.\cite{vuong2024estimands} 

Concerns have been raised about the applicability of estimands to evidence synthesis and (network) meta-analyses, by statisticians working in clinical development and the pre-market authorization (regulatory) setting,\cite{hedman2024estimand, russek2022discussion} but also by HTA bodies.\cite{morga2023intention} We recognize that many cautions remain, particularly regarding operational complexity and practical feasibility, e.g., in relation to HTA scoping or the analysis of secondary data of previously conducted trials, typically with limited input on primary data collection. As such, our proposal for the use of estimands is pragmatic in nature and does not necessarily align with the ``estimand thinking process'' of ICH E9 (R1). Our proposal also departs from the original idea by Remiro-Az\'ocar of replacing the PICO framework with `PICOSI'', incorporating all estimand attributes.\cite{remiro2022some} 

\begin{figure}[!htb]
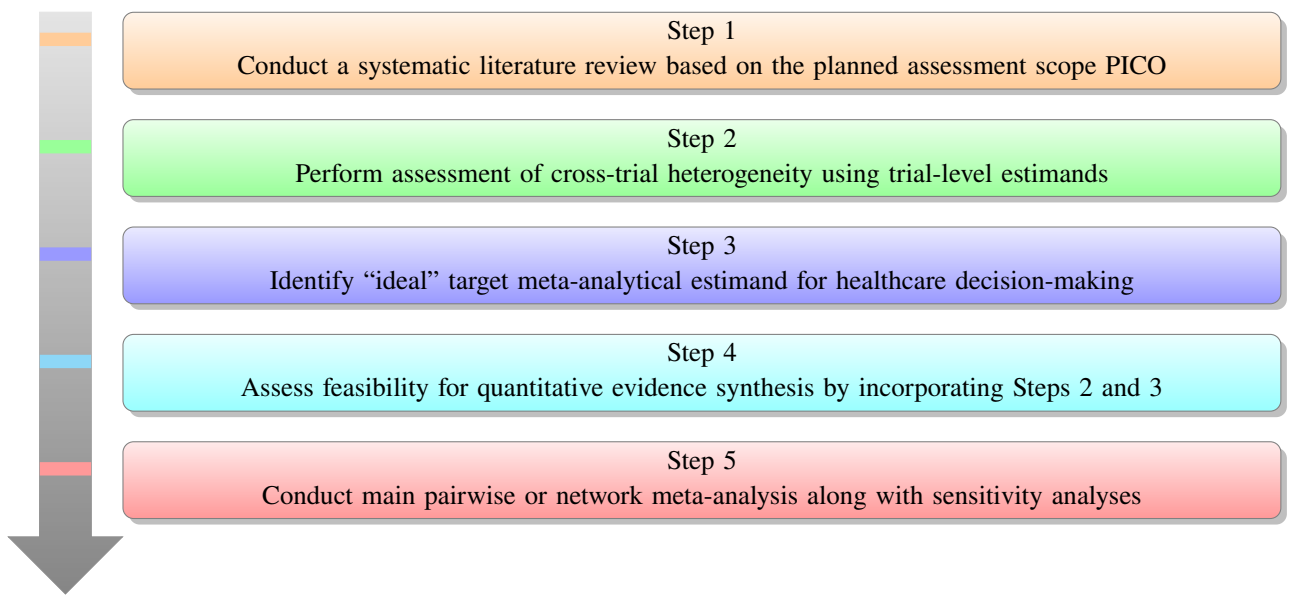

\centering
\smartdiagramx[priority descriptive diagram]{%
    Step 5\\Conduct main pairwise or network meta-analysis along with sensitivity analyses,
    Step 4\\Assess feasibility for quantitative evidence synthesis by incorporating Steps 2 and 3,
    Step 3\\Identify ``ideal'' target meta-analytical estimand for healthcare decision-making,
    Step 2\\Perform assessment of cross-trial heterogeneity using trial-level estimands,
    Step 1\\Conduct a systematic literature review based on the planned assessment scope PICO
}{}
\caption{Proposed roadmap for HTA evidence synthesis planning and execution, incorporating PICOs and estimands.}
\label{evisynth_Pipeline}  
\end{figure}

Our proposal could be integrated within a broader roadmap of evidence synthesis activities in HTA as follows (Figure \ref{evisynth_Pipeline}): (1) conduct a systematic literature review, with protocol and search terms based on the planned assessment scope PICO; (2) assess cross-trial heterogeneity using the trial-level estimands; (3) with stakeholder input, identify a suitable target meta-analytical estimand for healthcare decision-making; (4) perform a feasibility assessment for the quantitative evidence synthesis, based on Steps 2 and 3; and (5) conduct pairwise or network meta-analyses, perhaps with sensitivity analyses allowing for varying degrees of misalignment between the trial-level estimands. 

\smartdiagramset{
  descriptive items y sep = 4em,
  description font = \normalsize,
  description text width = 15cm,
}

\clearpage

Our proposal maintains the use of PICOs for qualitative assessments, while using estimands to assess the feasibility and relevance of quantitative analyses. Existing templates for meta-analytical feasibility assessments\cite{cope2014process} may require updating to reflect ICH E9 (R1), estimands and intercurrent event strategies. Importantly, trial-level estimands must be consistently recorded so that those performing systematic literature reviews can retrieve the relevant information from publications. Widespread adoption of estimands will expose the need for a more structured approach to the reporting and dissemination of trial results, e.g., via meta-data on treatment effect estimates and their associated estimand attributes on ClinicalTrials.gov. 

We acknowledge that the relative imprecision of PICOs is attractive to describe the broad ``totality'' of evidence in qualitative evidence syntheses, and to define the original scope of HTA decision problems. Nevertheless, this should not detract from the value of estimands in quantitative evidence syntheses. We advocate for the explicit integration of estimands into the planning of pairwise and network meta-analyses. One can formulate a PICO for scoping, then define different meta-analytical estimands. These can be used to restrict the original evidence base and perform more targeted (network) meta-analyses, precisely addressing the specific research question that might be of interest to healthcare decision-makers. The meta-analytical estimand can serve as a basis for cross-disciplinary discussion, to strengthen communication between different stakeholders about what a (network) meta-analysis aims to demonstrate, and to facilitate the interpretation of its results.\cite{lanius2025realizing} 

Moreover, the estimand framework provides a shared language to align trial design with both regulatory and HTA objectives.\cite{regulatory-HTAb} Insofar, HTA considerations have played a limited role in estimand selection, which has focused primarily on the perspective of key regulatory agencies. While these perspectives are important, HTA bodies may require different estimands for decision-making and views may differ between HTA agencies based on the context in which they make their decisions. Confirmatory clinical trials should be planned with both regulatory- and HTA-relevant estimands in mind.\cite{schiel2022commentary} Beyond traditional clinical trials, real-world data (RWD) are increasingly used to support HTA decision-making. Constructing appropriate estimands for the analysis of RWD -- and incorporating them into meta-analyses -- will involve specific considerations beyond those encountered for traditional clinical trials and remains a research priority. 

We encourage discussion and consensus-building among stakeholders within different therapeutic areas about which intercurrent events are most relevant, and which strategies are appropriate to handle them in specific decision-making contexts. In T2D, premature treatment discontinuation and the initiation of anti-diabetic rescue medication have been identified as key intercurrent events by regulatory stakeholders, with the European Medicines Agency publishing a therapeutic guideline (CPMP/EWP/1080/00 Rev.2) recommending a treatment policy strategy for handling premature treatment discontinuation and a hypothetical strategy for initiation of anti-diabetic rescue medication.\cite{EMAdiabetesguideline} Other therapeutic areas are yet to build consensus about intercurrent event strategies. In oncology, treatment switching is considered to be a major intercurrent event,\cite{morga2023intention} but current reporting practices to address treatment switching are inadequate, both in trial-level analyses and in meta-analyses.\cite{metcalfe2025treatment} Similar to the development of core patient characteristic sets and core outcome sets, defining important prognostic factors and outcomes to be measured and reported among specific therapeutic and disease settings, we would encourage stakeholder initiatives to define core intercurrent events and intercurrent event strategies.

Finally, we have focused on the role of intercurrent events and intercurrent event strategies in contextualizing the external validity of individual trial and meta-analytical results. Nevertheless, accounting for other estimand attributes is no less important when judging the transferability of results to any given clinical setting. For instance, target populations in RCTs are typically less heterogeneous and more narrowly defined than actual patient populations in routine clinical practice, partly to maximize statistical efficiency and power for hypothesis testing.\cite{ greenhouse2008generalizing} The external validity of RCTs can also be undermined if their treatment implementations differ from those in routine clinical practice, e.g., due to stringent prohibitions on the use of non-trial treatments, or if their endpoints have limited clinical relevance.\cite{rothwell2006factors} There remains a need to incentivize the use of estimands that prioritize real-world relevance and can provide the external validity that is necessary for actionable healthcare decision-making. 

\section*{Acknowledgments}

This article results from the work of the EIWG sub-team in HTA and RWE. We gratefully acknowledge the contributions of current and former members Amel Besseghir, Arthur Allignol, Barbara Rosettani, Frank Kleinjung and, especially, Anders Gorst-Rasmussen, who was instrumental in conceptualizing the article. Finally, we thank the Evidence Synthesis Scientific Working Group and the Estimands Scientific Working Group at Novo Nordisk for their very useful feedback.  

\subsection*{Competing interests statement}

Antonio Remiro-Az\'ocar is an employee of Novo Nordisk Pharma. Pepa Polavieja is an employee of Novo Nordisk Pharma and a minor shareholder in Eli Lilly. Emmanuelle Boutmy is an employee of Merck Healthcare KGaA and has minor stock ownership in Merck. Alessandro Ghiretti is an employee of Daiichi Sankyo Italia. Khadija Rerhou Rantell, David M. Phillippo and Jay J.H. Park declare no competing interests for this work. Lise Lotte Nystrup Husemoen, Helle Lynggaard and Robert Bauer are employees of Novo Nordisk A/S and hold shares in Novo Nordisk. Tatsiana Vaitsiakhovich is an employee of Sanofi-Aventis Deutschland GmbH and a minor shareholder in Bayer AG. Antonia Morga is an employee of Astellas Pharma Inc. 

\subsection*{Data availability statement}

The network meta-analyses are based on aggregate-level data, available in the SUSTAIN 7,\cite{pratley2018semaglutide} SUSTAIN FORTE\cite{frias2021efficacySEMA} and AWARD-11\cite{frias2021efficacy} publications. Code implementing the network meta-analyses is available at \url{https://github.com/antonio-ra/sema-estimands-NMA}. 

\subsection*{Funding statement}

David M. Phillippo is supported by the UK Medical Research Council (Grant Nos. MR/R025223/1 and MR/W016648/1).

% \end{itemize}

% {\fontsize{9pt}{10.8pt}\selectfont 

\bibliography{wileyNJD-AMA}

% }

%\nocite{*}% Show all bib entries - both cited and uncited; comment this line to view only cited bib entries;/

\end{document}